\documentclass[conference]{IEEEtran}
\IEEEoverridecommandlockouts

\usepackage{amsmath,amsfonts} 
\usepackage{graphicx}
\usepackage{textcomp}
\usepackage{array}
\usepackage{fancyhdr}
\usepackage[]{hyperref}
\usepackage{color}
\usepackage[table,dvipsnames]{xcolor}
\usepackage{fancybox} 
\usepackage{makecell}
\usepackage[export]{adjustbox}
\usepackage[ruled, vlined, linesnumbered]{algorithm2e}
\usepackage{algpseudocode}
\usepackage{multirow, booktabs}
\usepackage{tikz}
\usepackage{listings}
\usepackage{bbding}
\usepackage{pifont}
\usepackage{amsthm} 
\usepackage{colortbl}
\usepackage{tabularx}
\usepackage{enumitem} 
\usepackage[shortcuts]{extdash} 

\usepackage{marvosym}

\usepackage{soul}
\soulregister{\ref}7
\soulregister{\cite}7
\soulregister{\eqref}7

\def\BibTeX{{\rm B\kern-.05em{\sc i\kern-.025em b}\kern-.08em
    T\kern-.1667em\lower.7ex\hbox{E}\kern-.125emX}}

\def\sec{Section}

\def\eqn{Eq.}

\def\fig{Figure}

\def\tab{Table~}

\definecolor{keywordcolor}{rgb}{0.13,0.13,0.8}
\definecolor{commentcolor}{rgb}{0.13,0.5,0.13}
\definecolor{stringcolor}{rgb}{0.6,0.1,0.1}
\definecolor{funccolor}{rgb}{0.5,0,0.5}
\definecolor{emphcolor}{rgb}{0.8,0.0,0.0}

\lstdefinestyle{pytorch}{
    basicstyle=\ttfamily\small,
    breaklines=true,
    captionpos=b,
    escapechar=\%,         
    keepspaces=true,
    numbers=left,
    numbersep=8pt,
    showspaces=false,
    showstringspaces=false,
    showtabs=false,
    tabsize=2,
    language=Python,
    frame=single,           
    framerule=0.8pt,        
    framesep=6pt,           
    rulecolor=\color{gray}, 
    keywordstyle=\color{keywordcolor}\bfseries,
    commentstyle=\color{commentcolor},
    stringstyle=\color{stringcolor},
    morekeywords={nn, torch, Parameter, matmul, topk, bincount, relu},
    keywordstyle=[2]\color{funccolor}\bfseries, 
    keywordstyle=[3]\color{emphcolor}\bfseries, 
    morekeywords=[2]{torch, nn, matmul, topk, bincount, relu, matmul},
    backgroundcolor=,
}

\newcommand*\circled[1]{\tikz[baseline=(char.base)]{
            \node[shape=circle,fill, inner sep=0pt, minimum width=0.3cm] (char) {\textcolor{white}{#1}};}}
\newcommand*\whitecircled[1]{\tikz[baseline=(char.base)]{
            \node[shape=circle, draw=black, fill=white, line width=0.9 pt, inner sep=0=.1pt, minimum width=0.3cm] (char) {\textcolor{black}{#1}};}}

\definecolor{capcolor}{RGB}{51,51,204}

\newtoggle{todo}
\toggletrue{todo}

\newcommand{\update}[1]{\textcolor{black}{#1}}

\newtoggle{comment}
\togglefalse{comment} 

\SetAlgoSkip{smallskip}

\def\BibTeX{{\rm B\kern-.05em{\sc i\kern-.025em b}\kern-.08em
    T\kern-.1667em\lower.7ex\hbox{E}\kern-.125emX}}
\begin{document}

\title{Rethinking Unified Memory for NPU–PIM Systems: Dual-View Memory for Dynamic Inference of LLM}

\author{\IEEEauthorblockN{Shixin Zhao\textsuperscript{$\dagger$}}
\IEEEauthorblockA{\textit{Institute of Computing Technology} \\
\textit{Chinese Academy of Sciences}\\
\textit{University of Chinese Academy of Sciences
}\\
Beijing, China \\
zhaoshixin18@mails.ucas.ac.cn}
\and
\IEEEauthorblockN{Lian Liu\textsuperscript{$\dagger$}}
\IEEEauthorblockA{\textit{Institute of Computing Technology} \\
\textit{Chinese Academy of Sciences}\\
Beijing, China \\
liulian211@mails.ucas.ac.cn}
\and
\IEEEauthorblockN{Tianhua Han}
\IEEEauthorblockA{\textit{Institute of Computing Technology} \\
\textit{Chinese Academy of Sciences}\\
\textit{University of Chinese Academy of Sciences
}\\
Beijing, China \\
hantianhua22@mails.ucas.ac.cn}
\and
\IEEEauthorblockN{Mengdi Wang}
\IEEEauthorblockA{\textit{Institute of Computing Technology} \\
\textit{Chinese Academy of Sciences}\\
Beijing, China \\
wangmengdi@ict.ac.cn}
\and
\IEEEauthorblockN{Yinhe Han}
\IEEEauthorblockA{\textit{Institute of Computing Technology} \\
\textit{Chinese Academy of Sciences}\\
Beijing, China \\
yinhes@ict.ac.cn}
\and
\IEEEauthorblockN{Ying Wang\textsuperscript{\Letter}}
\IEEEauthorblockA{\textit{Institute of Computing Technology} \\
\textit{Chinese Academy of Sciences}\\
Beijing, China \\
wangying2009@ict.ac.cn}
}

\maketitle

\begin{abstract}
Heterogeneous architectures that combine neural processing unit (NPU) and processing-in-memory (PIM) are increasingly adopted to accelerate LLM inference. Prior work focuses on building a unified memory that allows NPUs and PIM to share data without duplication. However, these designs implicitly assume that each tensor is bound to a fixed execution device, and therefore rely on static, device-biased data mappings.
    
We observe that this assumption does not hold in modern LLM workloads. Due to phase changes (e.g., prefill vs. decode) and dynamic behaviors such as MoE routing, the optimal execution device for the same tensor can change at runtime. Under such dynamic execution, device-biased mappings become mismatched to access patterns, leading to \update{substantial bandwidth underutilization} and performance loss.
This paper presents PFM (PIM-as-Flexible-Memory), a dual-view memory system that decouples physical data layout from accessor-visible logical views. PFM stores data in a jointly optimized physical layout and exposes different logical interpretations to NPUs and PIM, enabling efficient access across devices without data duplication or relayout. We further design accessor-aware address translation and runtime scheduling mechanisms to support dynamic execution when LLM workloads fluctuate and the optimal execution device dynamically changes.
Our evaluation across LLMs shows that PFM improves end-to-end throughput by up to 2.32$\times$, demonstrating its effectiveness and broad applicability as a unified memory management solution for NPU-PIM systems.
\end{abstract}


\def\thefootnote{$\dagger$}\footnotetext{Both authors contributed equally to this research}\def\thefootnote{\arabic{footnote}}
\def\thefootnote{\Letter}\footnotetext{Corresponding author}\def\thefootnote{\arabic{footnote}}

 \section{Introduction}

Large language models (LLMs) power diverse applications, from natural language understanding~\cite{claude, chat-gpt, dubey2024llama3} to conversational AI~\cite{roziere2023code, xie2024waitgpt, kabakucs2024battle, chiang2023vicuna}. LLM inference is inherently heterogeneous: some operators are compute-intensive, whereas others are memory-bound~\cite{kwon2023efficient, hu2025lightllm, zhong2024distserve}. Consequently, NPU-PIM architectures are well-suited to LLM serving. NPUs accelerate compute-heavy operators with high throughput, while PIM offloads memory-bound operators through near-data execution~\cite{kim2024sk, wu2024pim, li2025h2, yun2024duplex}.

However, realizing this potential requires one prerequisite: NPUs and PIM must efficiently share the same data. Existing designs still fail to provide such support for modern LLM inference because they are fundamentally built for \emph{static} execution, whereas LLM inference is increasingly \emph{dynamic}. In practice, the same tensor may be accessed by different \emph{accessors} (i.e., NPUs or PIM) at different times since the best execution device for the same operator varies~\cite{zhong2024distserve,fang2025klotski,cao2025moe}. Once this happens, existing systems either incur costly migration or suffer \update{large bandwidth loss} because the data layout is optimized for a mismatched accessor.


This challenge arises from the mismatch between the data layout preferences of NPUs and PIM. NPUs prefer fine-grained \textit{channel interleaving} to maximize bus-level bandwidth by spreading consecutive accesses across channels~\cite{pessl2016drama}. PIM, in contrast, prefers \textit{bank-level parallelism} together with \textit{intra-bank data continuity} to reduce inter-bank traffic during near-data execution~\cite{zhao2024pim, seo2025facil}. As a result, a layout optimized only for one device is often inefficient for the other.
A straightforward way to accommodate these conflicting requirements is to physically separate the memory spaces of NPU and PIM. Such \textbf{separated memory systems} (e.g., UPMEM and HBM-PIM designs) preserve each device's preferred layout~\cite{devaux2019true, lee20221ynm}. However, whenever data must be reused across devices, they incur explicit migration or duplication, leading to substantial latency and capacity overhead.
To eliminate these costs, recent work has moved toward \textbf{unified memory systems}, where NPU and PIM share one physical address space through a layout-aware memory controller. Representative examples include IANUS~\cite{seo2024ianus}, which adopts a globally PIM-friendly organization, and FACIL~\cite{seo2025facil}, which supports multiple address mappings within a single memory space and pre-execution controller-level mapping selection. This shift is an important step forward because it removes explicit physical partitioning and makes cross-device sharing easier.

\begin{figure}
    \centering
    \includegraphics[width=\linewidth]{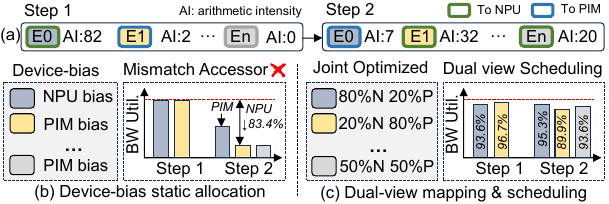}
    \caption{Motivation for PFM. (a) Step-varying expert intensity in MoE decoding changes the preferred execution device. (b) Existing device-biased mappings cause large bandwidth loss under device switching. (c) PFM sustains efficient NPU--PIM execution via dual-view mapping and runtime scheduling when the accessor changes.}
    \label{fig:intro}
    \vspace{-\baselineskip}
\end{figure}

Despite these advances, existing unified memory architectures remain insufficient for the demands of dynamic LLM inference. These designs share a fundamental assumption: \textit{each tensor is effectively bound to a fixed execution device at allocation time and therefore mapped according to a static, device-biased layout}. This assumption no longer holds in modern LLM inference workloads, as illustrated in \fig~\ref{fig:intro}(a). For example, in Mixture-of-Experts (MoE) models, dynamic routing causes expert activation and arithmetic intensity to fluctuate across decoding steps. 
Under such dynamic access, \textit{the optimal execution device for the same tensor can change over time}, and existing static device-biased mappings become a \update{performance bottleneck}. As illustrated in \fig~\ref{fig:intro}(b), when the runtime accessor mismatches the layout bias, effective bandwidth can drop to less than 20\% of peak. \update{Fig.~\ref{fig:intro} is based on profiled MoE execution and tensor-access replay under our evaluation setup, with device-biased mappings following prior unified-memory designs~\cite{seo2025facil}.} Even FACIL, the most flexible prior unified memory design, achieves only 29.5\% of the performance of an oracle that dynamically selects both execution device and layout. These results reveal that the true bottleneck is not physical separation itself, but the inability of existing unified memory systems to support \emph{dynamic cross-device access}. A natural solution is to relayout data at runtime whenever the preferred accessor changes. However, this is impractical for LLM inference, where execution preference can vary across decoding steps, making frequent relayout \update{high-overhead}.



\textbf{Key Insight.} Different from prior unified memory designs, efficient NPU-PIM memory management for dynamic LLM inference should \emph{decouple physical layout from accessor-visible interpretation}. Rather than physically duplicating data or repeatedly relayouts when execution shifts between NPU and PIM, a single physical layout can be jointly optimized once and exposed through different logical views tailored to each accessor. This enables both devices to access the same data efficiently without migration overhead.



Based on this insight, we propose \textbf{PFM} (PIM-as-Flexible-Memory), a lightweight hardware-software co-design that turns PIM into unified, flexible memory for NPU-PIM systems. As illustrated in \fig~\ref{fig:intro}(c), PFM maintains one continuous address space while exposing device-optimized logical views, thereby enabling data sharing without duplication and efficient execution migration between NPU and PIM. 
PFM is built around two key mechanisms:

\update{(1) \textbf{Dual-view mapping optimization.}}
PFM manages LLM-critical data (e.g., KV caches and expert weights) at \emph{superpage} granularity\footnote{We use the term \textit{superpage} to refer to large pages~\cite{vespa_superpages, usenix_atc20_superpage}.} and optimizes fine-grained placement \emph{within} each superpage~\cite{kwon2023efficient, vespa_superpages, usenix_atc20_superpage}. Using offline profiling, PFM formulates physical-address-to-hardware-address mapping as an optimization problem that balances the NPU's demand for multi-channel interleaving with the PIM's preference for bank-local continuity and parallelism. The result is an operator-specific superpage layout that supports both devices efficiently.

\update{(2) \textbf{Dual-view addressing \& scheduling for NPU-PIM.}} At runtime, PFM sustains high bandwidth through two lightweight extensions to the memory controller: 1) an \textbf{Address Remapping Unit (ARU)} that translates physical addresses into hardware addresses by applying the precomputed operator-specific mappings on the fly; and 2) a \textbf{Flexible Access Scheduler (FAS)} that enforces NPU-oriented and PIM-oriented scheduling policies on their respective request streams and arbitrates residual conflicts to maximize effective bandwidth.


Our main contributions are as follows:
\begin{itemize}
    \item We identify dynamic execution-device switching as a key limitation of existing unified memory systems for NPU–PIM architectures, and show that static device-biased mappings lead to \update{substantial performance degradation}. 
    \item We present PFM, the first flexible memory management that combines offline dual-view mapping optimization with runtime dual-view addressing \& scheduling to decouple physical layout from accessor-visible logical views, enabling efficient cross-device execution without data duplication.
    \item We integrate PFM into mainstream LLM frameworks and demonstrate up to 2.32$\times$ throughput improvement on representative LLMs over the best prior unified memory design.

\end{itemize}


  \section{Background}
\begin{figure}
    \centering
    \includegraphics[width=0.9\linewidth]{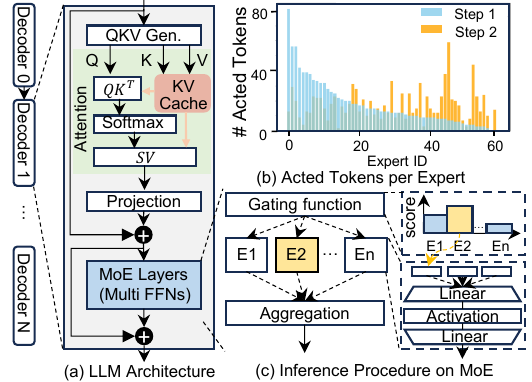}
    \caption{LLM architecture and procedure. \textbf{(a)} Classic LLM
    architecture. \textbf{(b)} Dynamic expert activation behaviors during the
    decode step in the MoE model.\textbf{ (c)} The architecture of MoE layers with
    the example of expert selection.}
    \label{fig:LLMarchi}
    \vspace{-\baselineskip}
\end{figure}

\subsection{LLM Architecture \& Procedure}
\label{sec:llmbackground}

\subsubsection{LLM Architecture}
As shown in \fig~\ref{fig:LLMarchi}(a), modern LLMs consist of stacked
Transformer blocks~\cite{workshop2022bloom, chowdhery2023palm}, each combining a
self-attention layer for long-range dependency modeling and a feed-forward network
(FFN) for nonlinear transformation. To scale model capacity, recent models adopt
the mixture-of-experts (MoE) architecture~\cite{jiang2024mixtral, liu2024deepseek},
where the FFN is replaced by multiple parallel expert networks, as illustrated in
\fig~\ref{fig:LLMarchi}(c). During inference, only a subset of experts is
activated per token, enabling massive parameter counts without proportional computational
cost~\cite{fedus2022switch, zhou2022mixture}.

\subsubsection{LLM Inference Procedure}
LLM inference comprises two phases: the \textit{prefill} phase for prompt processing and the autoregressive \textit{decode} phase for token generation. Each phase consists of several core operators, primarily self-attention and FFN or MoE layers. These phases and operators exhibit distinct computational characteristics, introducing variability at both coarse and fine granularities. \textbf{At the phase level}, the \textit{prefill} phase is compute-intensive, while the \textit{decode} phase is memory-bound due to the autoregressive decoding manner and frequent KV cache accesses~\cite{kwon2023efficient, patel2024splitwise}. \textbf{At the operator level}, MoE architectures introduce finer-grained variability. As illustrated in \fig~\ref{fig:LLMarchi}(b), dynamic expert selection leads to fluctuations in arithmetic intensity across steps, shifting between compute-intensive and memory-bound patterns. This multi-dimensional variability demands an acceleration system that adapts its compute and memory resources in real-time to maintain high efficiency.

\subsection{NPU-PIM System for LLM Inference}
Heterogeneous NPU-PIM architectures~\cite{he2025papi, seo2024ianus, li2025h2} support efficient LLM inference by leveraging complementary strengths of different processing units. In these systems, LLM operators are partitioned based on their computational characteristics: the NPU handles compute-intensive kernels (e.g., GEMM in the prefill phase), while PIM modules are dedicated to memory-bound operations (e.g., decode-time attention~\cite{park2024attacc, heo2024neupims} or MoE experts with low arithmetic intensity~\cite{yun2024duplex, kim2024monde}). Compared to NPU-only systems, this collaborative execution achieves up to 3.0$\times$ speedup~\cite{yun2024duplex, heo2024neupims, he2025papi}.

\begin{figure}
    \centering
    \includegraphics[width=\linewidth]{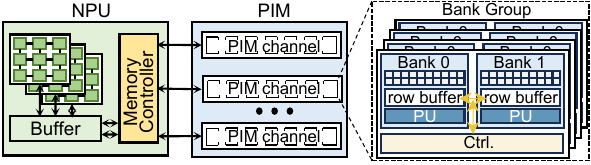}
    \caption{Heterogeneous NPU-PIM architecture.}
    \label{fig:hybridarchi}
    \vspace{-\baselineskip}
\end{figure}

As illustrated in \fig~\ref{fig:hybridarchi}(a), the architecture integrates a high-performance NPU with PIM-enabled memory modules via a high-throughput interconnect. The NPU features compute units such as tensor cores, while PIM integrates lightweight processing units (PUs) near DRAM arrays to exploit massive internal bandwidth. In such a system, PIM-enabled modules are an integral part of the NPU memory hierarchy, sharing the same physical memory space.

\subsection{NPU- v.s. PIM-optimized Memory Layout}

The physical realization of NPU-PIM heterogeneous collaboration requires a deep understanding of their divergent data layout requirements. NPUs and PIM units constrain data mapping to the physical DRAM hierarchy in distinct ways to maximize their respective efficiencies.

\begin{figure}[t]
    \centering
    \includegraphics[width=\linewidth]{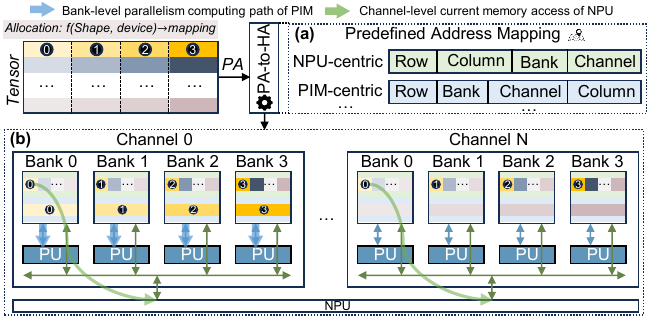}
    \caption{\update{Address mapping examples for NPU- and PIM-optimized layouts. (a) Address-bit placement for channel-interleaved NPU access and bank-localized PIM access. (b) Resulting data layouts, where green denotes the NPU-optimized layout and blue denotes the PIM-optimized layout.}}
    \label{fig:mapping}
    \vspace{-\baselineskip}
\end{figure}

NPUs favor \textbf{channel-interleaved} layouts to support contiguous, vectorized
accesses. As the green line in \fig~\ref{fig:mapping} shows, consecutive elements within
one data block (e.g., \circled{0}) are partitioned across different channels, while
distinct data blocks are allocated to different banks. This mapping strategy is
adopted because it enables the memory controller to activate multiple channels
concurrently, thereby maximizing bus-level bandwidth utilization. Moreover,
within each channel, elements are preferentially placed in the same row to
minimize row activation overhead.

In contrast, PIM units prioritize \textbf{bank-level parallelism} while requiring
\textbf{intra-bank data continuity} for local processing. As the blue line in \fig~\ref{fig:mapping} shows,
PIM-optimized layouts map contiguous data blocks within individual banks (e.g. \circled{0} entire in channel 0, bank 0), rather
than fragmenting them across banks and channels. This organization ensures
each PIM unit has sufficient local data for computation while enabling a single
instruction to trigger parallel in-situ execution across all banks within a row,
maximizing overall compute throughput. Due to these divergent layout requirements,
prior PIM-based systems often rely on a \textbf{separated memory management}
where the NPU and PIM operate in isolated physical regions~\cite{upmem_sdk, heo2024neupims}. However, this necessitates explicit data migration and layout transformation, incurring
substantial performance penalties and memory capacity overhead.

\subsection{Unified Memory Management on NPU-PIM}

To address these inefficiencies, recent research has pursued \textbf{unified memory management}, enabling the NPU and PIM to share a physical memory space. The cornerstone of these unified systems is a layout-aware memory controller (MC) that supports multiple DRAM address mapping schemes concurrently.

Specifically, existing unified memory architectures manage layout diversity through either a globally uniform layout (e.g., IANUS~\cite{seo2024ianus} applying a universal PIM-friendly mapping) or static, device-specific layouts~\cite{seo2025facil, park2024attacc}. FACIL~\cite{seo2025facil}, a typical device-specific unified memory management strategy, relies on static allocation and device-biased address mapping, binding tensors to predefined configurations (e.g., NPU- or PIM-centric) at compile time. During each memory access, the memory controller retrieves these strategies to translate physical addresses into DRAM-specific indices. Consequently, existing systems assume the target execution device for an operator remains constant at runtime, requiring device-biased mapping decisions.

\begin{figure}
    \centering
    \includegraphics[width=\linewidth]{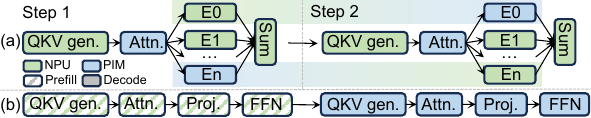}
    \caption{LLM inference on NPU-PIM system.}
    \label{fig:llmscheduling}
    \vspace{-\baselineskip}
\end{figure}
  \section{Limitations \& Motivation }\label{sec:motivation}

\subsection{Runtime Dynamics Requirements for LLM}
\textbf{Runtime Dynamism in LLM Inference.} LLM inference workloads exhibit runtime dynamism. In Mixture-of-Experts (MoE) architectures, expert activation patterns and arithmetic intensity fluctuate across tokens and decoding steps~\cite{yu2025orderschaosenhancinglargescale}. As shown in \fig\ref{fig:LLMarchi}(b), in Qwen1.5-MoE-A2.7B~\cite{qwen2025qwen15moea27b}, expert 0's demand pivots from highly compute-intensive (arithmetic intensity of 78) at step 0 to \update{memory-bound} (arithmetic intensity of 6) at step 1. This implies the optimal device for a given operator changes dynamically. As shown in \fig~\ref{fig:llmscheduling}, the same expert can be dispatched to different devices across adjacent decoding steps. Beyond MoE layers, other operators exhibit similar dynamic preferences. In small-batch scenarios, attention's QKV generation demands high compute throughput during prefill but becomes memory-bound during decode, favoring different devices. Existing unified memory systems~\cite{seo2025facil, seo2024ianus} assume a tensor's preferred device can be statically determined at allocation time, leading to device-biased address mappings that cannot adapt to runtime shifts.

\begin{figure}
    \centering
    \includegraphics[width=0.9\linewidth]{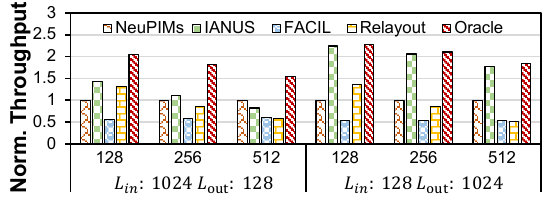}
    \caption{\update{Normalized end-to-end inference throughput (tokens/s) on GPT-OSS 120B~\cite{gpt-oss} using real-world expert activation traces from LMSYS-Chat-1M~\cite{zheng2023lmsyschat1m}. }}
    \label{fig:motivate-throughput}
    \vspace{-\baselineskip}
\end{figure}
\textbf{Quantitative Impact Analysis.} To quantify the impact of this static allocation, \fig~\ref{fig:motivate-throughput} compares the end-to-end inference throughput of GPT-OSS 120B~\cite{gpt-oss} under diverse representative strategies. \update{NeuPIMs represents separated memory management, binding attention/KV-cache-dominated execution to PIM and FFN execution to NPU. IANUS uses PIM-optimized mapping for all operators. FACIL, representing existing unified memory systems, enables runtime operator migration but retains device-biased address mappings. Attention tensors use PIM-optimized mappings; for MoE experts, half use NPU-optimized mappings while the others use PIM-optimized mappings after prefilling; all other tensors use NPU-optimized mappings. Oracle uses the same NPU-PIM hardware as the evaluated systems but assumes ideal runtime operator placement and ideal layout support: each operator is executed on the better device and fully exploits that device's bandwidth/compute capability without scheduling or relayout overhead. All strategies are evaluated by replaying the corresponding operator/tensor access patterns on our HBM-PIM memory model, using the same hardware configuration, timing parameters, and PIM organization as Section~\ref{sec:experimental-setup}.}

\begin{figure}
    \centering
    \includegraphics[width=\linewidth]{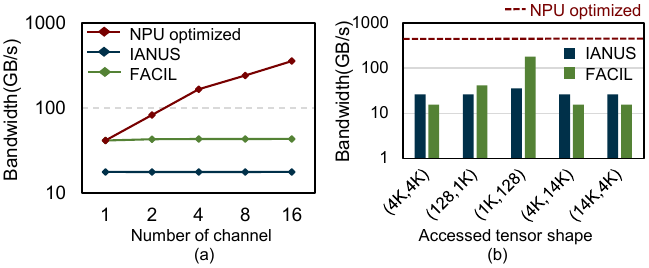}
    \caption{The bandwidth under different mapping strategies when accessing an HBM2e stack. \textbf{(a)} Achievable bandwidth under varying numbers of channels; for the same number, bandwidth is affected by row conflicts. \textbf{(b)} Achievable bandwidth under the address mappings used in FACIL~\cite{seo2025facil} and IANUS~\cite{seo2024ianus} when loading the weight/KV tensor of Mixtral~\cite{jiang2024mixtral}.}
    \label{fig:bandwidth}
\vspace{-\baselineskip}
\end{figure}

The Oracle design achieves a speedup of up to 2.26$\times$ over the NeuPIMs strategy, while FACIL captures only 29.5\% of this potential on average. This result highlights the substantial potential of runtime device switching for LLM inference, if data can remain efficiently accessible after migration. However, the performance gap of FACIL reveals a fundamental limitation: despite providing a unified address space and supporting migration, current designs fail to resolve the underlying data layout mismatch during runtime. The root cause is that a device-biased mapping becomes \update{inefficient} when the accessor changes. As shown in \fig~\ref{fig:bandwidth}, when an operator migrates to a device that does not match its device-biased mapping (e.g., NPU accessing a PIM-optimized layout), bandwidth degradation limits execution efficiency to less than 20\% of peak.

\subsection{From Static Bias to Joint Potential}

To bridge the performance gap, an ideal system should adaptively optimize data access based on the runtime execution device. We analyze two potential approaches: runtime relayout and joint mapping optimization.

\update{\textbf{The Overhead of Runtime Relayout.} One approach is to perform dynamic relayout whenever an operator is scheduled to a different device. We compare a dynamic \textbf{Relayout} strategy against Oracle. Relayout periodically remaps expert weights to the NPU- or PIM-preferred layout according to recent expert-execution history, and we sweep relayout intervals of 1, 5, 10, and 20 decoding steps and report the best end-to-end result for each model. Even with this optimistic interval selection and considering only read-out/write-back data movement, Relayout remains 13.61\% below Oracle on DeepSeekMoE, 53.6\% below Oracle on GPT-OSS, and 49.83\% below Oracle on Mixtral, while relayout traffic accounts for 10.13\%, 46.32\%, and 42.95\% of end-to-end execution time on average, respectively. As shown in \fig~\ref{fig:motivate-throughput}, Relayout can even underperform a single device-biased strategy because the data movement overhead offsets the benefit of switching to a more favorable layout. Therefore, for MoE-scale LLMs with step-level expert-activation changes, runtime relayout is not a practical strategy.}

\textbf{The Potential for Joint Optimization.} A promising direction is to find a \textit{joint address mapping} that satisfies both NPU and PIM requirements without physical relayout. Modern large-page management (e.g., 2 MB superpage) provides the physical foundation. For instance, in an HBM3 stack (16 channels, 1024 banks) with a 1 KB row size, a single 2 MB page spans all 1024 banks with 2 KB per bank. This allows the NPU to exploit multi-channel interleaving while ensuring each PIM unit with bank-level parallelism.
Achieving peak performance requires the memory controller to dynamically schedule DRAM accesses to match the accessor patterns, a capability missing in current device-biased unified memory systems. For NPU accesses, the controller must "stitch" data from different channels correctly; for PIM, it must synchronize commands across parallel banks. This dual-access scheduling challenge, combined with the need to search for optimal joint mappings, motivates our flexible and accessor-aware design.

\subsection{Key Insight: Flexible Logical View}

These observations reveal that the fundamental bottleneck is not the lack of a shared physical space, but the \textit{rigidity} of the logical-to-physical mapping. We propose a flexible memory management architecture that decouples the \textbf{logical data view} from the \textbf{physical DRAM mapping}. Our key insight is to allow a single, jointly-optimized physical layout to be perceived differently depending on the accessor. By employing a balanced physical mapping that spans the DRAM hierarchy, combined with an \textbf{accessor-aware memory controller} that performs runtime address translation and request scheduling, we simultaneously provide high-bandwidth, channel-interleaved access for the NPU and bank-parallel execution for PIM. \update{This logical-view flexibility avoids repeated data relayout while allowing the runtime scheduler to adapt operator placement to changing arithmetic intensity.}

  \section{PFM: PIM as Flexible Memory}\label{sec:pfm-method}
We propose PFM (PIM-as-Flexible-Memory), a lightweight hardware-software co-design that unifies PIM and NPU memory spaces while preserving device-optimal data layouts. This section outlines PFM’s principles and interface (\sec~\ref{sec:overview}), presents its dual-view mapping (\sec~\ref{sec:dual-mode mapping}), enabling simultaneous, access-specific views, and details the runtime address translation and scheduling mechanisms that make unified, flexible memory practical (\sec~\ref{sec:memory-aware scheduling}).

\subsection{Overview}\label{sec:overview}

\subsubsection{Design Principles}
PFM adheres to two guiding principles: \textbf{minimal hardware modification} and \textbf{maximal software compatibility}. On the software side, a lightweight user-level API provides efficient management over memory allocation and automatic layout optimization. Computing kernels (e.g., cuBLAS~\cite{cublas}) remain unchanged. On the hardware side, modifications are confined to the memory controller (named as the PFM controller), which supports dual-view addressing and flexible access scheduling. We do not modify DRAM cell arrays, timing protocols, or NPU cores.

\begin{figure}
    \centering
    \includegraphics[width=0.98\linewidth]{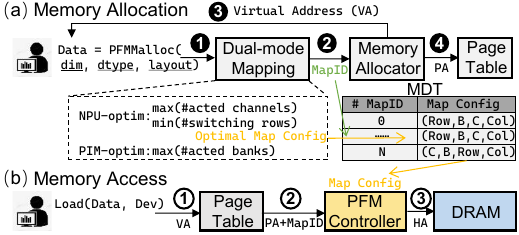}
    \caption{Memory allocation and access with PFM.}
    \label{fig:pfm-malloc}
\vspace{-\baselineskip}
\end{figure}


\subsubsection{PFM Interface for Memory Allocation and Access}
Guided by these principles, PFM introduces \texttt{PFMMalloc}, which couples allocation with layout optimization. As depicted in \fig~\ref{fig:pfm-malloc}(a), it first \circled{1} analyzes tensor dimensions, data types, and logical layout (e.g., column- or row-major), then models access performance for both NPU and PIM. By solving a multi-objective optimization problem, PFM \circled{2} derives the optimal PA-to-HA (Physical Address to Hardware Address) mapping. This mapping is stored in the PFM controller's Mapping Descriptor Table (MDT), enabling a “dual-optimized” abstraction. During runtime access, PFM leverages DRAM command scheduling and lightweight data relayout to optimize the access efficiency. It provides NPU with cross-channel contiguous access and PIM with bank-aligned distribution. Then \circled{3} the virtual address (VA) is returned to the user, and \circled{4} the physical address (PA) is assigned to the page table. 

\update{Based on the MapID generated above, PFMMalloc preserves conventional allocation semantics while augmenting each allocated superpage with mapping metadata. After physical superpages are assigned, PFMMalloc records the MapID and the corresponding physical footprint class, which captures the channels, banks, and rows occupied by that mapping. During deallocation, PFMMalloc removes the MapID binding and returns each superpage to the free list of its footprint class. A later allocation can reuse that space whenever its required footprint is compatible, even if the new tensor uses a different MapID.}

\update{For example, two MapIDs may swap the channel and row bit positions, e.g., using the lower three bits for channels versus rows. Although their access orders differ, both can occupy the same channel-row footprint, so a freed superpage from one MapID can be reused by the other. Thus, PFM avoids maintaining isolated memory pools for each MapID.}

\update{If no compatible free superpage is available, PFMMalloc falls back to allocating a new superpage and binding it to the requested MapID; correctness does not depend on reusing the same physical location. This fallback may reduce reuse opportunities under severe fragmentation, but it preserves the same VA-to-PA semantics and avoids assuming ideal placement. Similarly, in KV-cache eviction and restoration, the runtime can still follow block-level mechanisms such as PagedAttention: an evicted block releases its backing superpages, and a restored block only needs compatible superpages with the proper MapID, not the exact same physical addresses.}

\emph{Access path.} As illustrated in \fig~\ref{fig:pfm-malloc}(b), memory requests first undergo standard VA-to-PA translation \whitecircled{1} according to the page table. Then, it \whitecircled{2} uses the MDT and the request \textbf{origin} (NPU or PIM) to resolve hardware addresses and \whitecircled{3} applies a \textbf{dual-view memory scheduler}: NPU requests are reordered for channel concurrency and row locality; PIM requests are batched for bank-parallel broadcasts. The scheduler in the PFM controller arbitrates between the two modes, prioritizing latency-sensitive requests while enhancing overall efficiency through batching and exploiting idle periods.

\subsection{Dual-view Mapping Optimization}\label{sec:dual-mode mapping}


This section details the dual-view mapping optimization, which aims to identify joint address mappings that maximize end-to-end performance. We employ a fast analytical model to accurately compare relative access efficiencies between NPU and PIM, enabling rapid mapping search.

\subsubsection{Page Size}
PFM uses 2 MB superpages for co-optimized mapping as they : (1) match existing NPU/GPU huge page interface~\cite{a100paper,h100paper}; (2) reduce TLB pressure for large-scale LLM weight/KV caches; (3) provide sufficient layout optimization space. The expanded intra-page address space enables finer control over data placement across channels and banks, allowing PFM to simultaneously satisfy the NPU’s and the PIM’s demand.

\begin{figure}
    \centering
    \includegraphics[width=\linewidth]{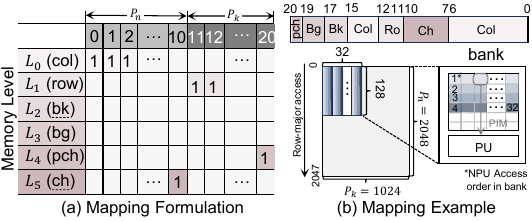}
    \caption{Illustration of address mapping formulation.}
    \label{fig:formulation}
\vspace{-\baselineskip}
\end{figure}

\subsubsection{Mapping Formulation}
Consider a two-dimensional $(N, K)$ tensor in a given layout (e.g., column-major) placed in memory with levels ${L_0, L_1, \dots, L_{m-1}}$, including channel, bank, row, and column, etc. The memory space is divided into contiguous 2 MB pages. Within each page, PFM solves a configurable mapping $f: \text{logical index} \rightarrow \text{hardware address fields}$, which determines the PA-to-HA mapping. As shown in \fig~\ref{fig:formulation}(a), the mapping can be represented by a binary matrix $\mathbf{X} \in \{0,1\}^{m \times n}$, where $n$ is the number of address bits (e.g., 21 bits for a 2 MB page). Each column $j$ corresponds to an address bit, and each row $i$ to a memory level $L_i$, with $\mathbf{X}_{i,j} = 1$ indicating that address bit $j$ is assigned to level $L_i$. The value of $n$ is expressed as $n = p_n + p_k$, where $p_n$ and $p_k$ are determined by the tensor’s layout. For example, under a column-major layout with the tensor size $(N, K)$, we have $p_n = \log_2(N)$ and $p_k = \log_2\left(\frac{\text{page\_size}}{N}\right)$, so that the lower $p_n$ bits of the address encode the contiguous dimension, while the higher $p_k$ bits represent the stride dimension. In contrast, for a row-major layout, we have $p_n = \log_2\left(\frac{\text{page\_size}}{K}\right)$ and $p_k = \log_2(K)$, with the lower $p_k$ bits now representing the contiguous dimension. This ensures that the mapping accurately reflects the memory access patterns of different layouts.
\update{Meanwhile, the two-dimensional notation is an abstraction of the linearized memory layout rather than a restriction to matrix weights. Higher-dimensional tensors are first linearized according to the framework layout, while preserving the contiguous dimension that dominates memory access. For example, a PagedAttention-style KV-cache block with shape $[\text{block\_size}, \text{num\_kv\_heads}, \text{head\_dim}]$ is modeled as $[\text{block\_size},$ $ \text{num\_kv\_heads} \times \text{head\_dim}]$: the head dimension within each token is treated as the contiguous dimension, and the block dimension captures the token sequence within the page. The resulting $(N,K)$ representation is then handled by the same mapping formulation.}

\noindent\textit{\textbf{Constraints.}}
Based on the above formalism, we define two key constraints to satisfy the validity of mapping.

\textbf{Capacity constraints.} Each memory level $L_i$ can address only a finite number of locations, determined by its physical capacity. We denote the address bitwidth available for $L_i$ as $B_i$, representing the number of bits required to uniquely identify each addressable location. For instance, the number of banks within each bank group typically equals to 4, so that the address bits that can be allocated for bank level is 2.
When assigning address bits to a memory level, the number of bits allocated must not exceed $B_i$:
$
\sum_{j=1}^{n} X_{i,j} \leq B_i.
$

\textbf{Structural constraints.} Each physical address dimension must be mapped to exactly one memory level. For every address bit $j$,  
$
\sum_{i=1}^{m} X_{i,j} = 1.
$  
This enforces a one-to-one assignment of address bits to memory levels.


\subsubsection{Optimization Goal}
Given the formulation, PFM defines performance goals for NPU and PIM, respectively.

\begin{figure*}[t]
    \centering
    \includegraphics[width=\linewidth]{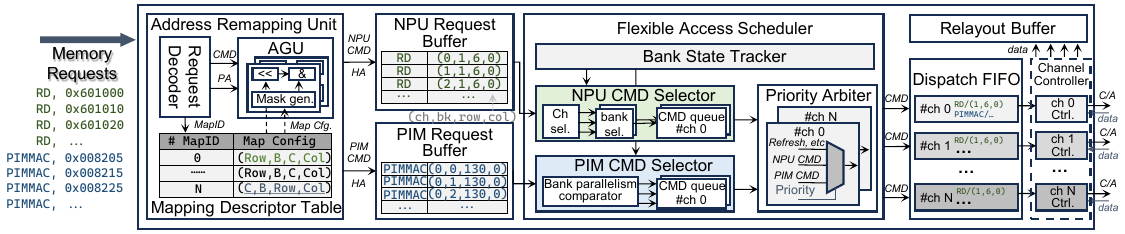}
    \caption{The PFM controller path: dual-view hardware address translation and scheduling.}
    \label{fig:scheduler}
\vspace{-\baselineskip}
\end{figure*}

\textbf{NPU-objective.} 
NPU favors spreading across channels with minimal row thrashing. Thus, we define the bandwidth utilization for NPU as:
\begin{equation}
    \text{NPU}(f) = \frac{\mathcal{B}_{\text{NPU}}}{\mathcal{B}_{\text{peak}}}
\end{equation}
where $\mathcal{B}_{\text{NPU}}$ is the achievable bandwidth under mapping $f$, and $\mathcal{B}_{\text{peak}}$ is the theoretical peak bandwidth.

The achievable bandwidth is determined by: $\mathcal{B}_{\text{NPU}} = \frac{\text{data\_size}}{T_{\text{mem}}(f)}$, where $T_{\text{mem}}(f)$ is the time to transfer ($p_n,p_k$) data under given mapping $f$. This is governed by the slowest parallel unit in the memory levels: 
\begin{equation}
    T_{\text{mem}}(f) = \max_{l \in L_{\text{parallel}}} T_{\text{access}}(l,f)
\label{eq:npu-parallel}
\end{equation}
where $L_{\text{parallel}}$ denotes levels that support concurrent access, such as independent pseudo-channels (PCHs) in HBM. Within each PCH, lower memory levels (e.g., bank groups, banks) operate serially. Therefore, $T_{\text{mem}}(f) = \max_{i \in \{\text{pch}\}} T_{\text{access}}(i,f)$, where $T_{\text{access}}(\text{pch},f)$ aggregates over its bank groups and banks accordingly. For a given bank, access time depends on the times of row and column activations:
\begin{equation}
T_{\text{access}}(\text{bank}, f) = \#_{\text{row-act}} \cdot t_{\text{RCD}} + \#_{\text{col}} \cdot t_{\text{CCD}} 
\label{eq:bank-time}
\end{equation}
where $t_{\text{RCD}}$ and $t_{\text{CCD}}$ are DRAM row activation and column-to-column delay. \update{In the mapping solver, $t_{\text{CCD}}$ is selected according to the bank-group relationship of consecutive column commands: accesses to different bank groups use $t_{CCD\_S}$, while accesses within the same bank group use $t_{CCD\_L}$. Thus, the analytical search accounts for bank-group interleaving, and the final cycle-level evaluation further enforces all DRAM timing constraints through Ramulator.} The row activation times $\#_{\text{row-act}}$ and column access times $\#_{\text{col}}$ are determined jointly by the mapping matrix $\mathbf{X}$ and the tensor’s layout. For example, here we let $p_n$ and $p_k$ denote the number of address bits associated with the contiguous and stride dimensions, respectively.
As illustrated in \fig~\ref{fig:formulation}(b), one row activation occurs whenever the accessed address maps to a different row within the same bank. The total times of row activation can be expressed as:
\begin{equation}
    \#_{\text{row-act}} = 2^{\sum_{j=0}^{p_n-1}\mathbf{X}_{row,j}} \cdot 2^{\sum_{j=p_n}^n\mathbf{X}_{col,j}},
\label{eq:row-activation}
\end{equation}
where $\sum_{j=0}^{p_n-1}\mathbf{X}_{row,j}$ is the number of contiguous-dimension bits mapped to the row field, capturing the distinct rows visited along the contiguous axis; $\sum_{j=p_n}^n\mathbf{X}_{col,j}$ is the number of stride-dimension bits mapped to the column field, which causes the same row to be revisited due to non-contiguous accesses.
The number of column access is given by:$\#_{\text{col}}=2^{\sum_j\mathbf{X}_{col,j}}$.
This formulation explicitly links access latency to both the mapping function $f$ and the tensor’s logical layout ($p_n$, $p_k$), enabling accurate performance estimation across different memory mappings.

Furthermore, the theoretical peak bandwidth is computed as:
$\mathcal{B}_{\text{peak}} = (\text{\#channels} \times \text{\#pins per channel}) \times \text{frequency}$.

\textbf{PIM-objective.}
PIM favors (i) many banks in parallel and (ii) low intra-bank serialization by activating more banks per batch while minimizing per-bank access latency. We capture this trade-off with the bank-level parallelism gain:
\begin{equation}
    \text{PIM}(f) = \frac{\#_\text{active\_banks}(f)}{\#\text{banks}_{\text{total}}} \cdot \frac{T_{\text{ref}}}{T_{\text{bank}}(f)} 
\label{eq:bank-gain}
\end{equation}
where the first term $\#_\text{active\_banks}(f) / \#\text{banks}_{\text{total}}$ measures the fraction of total banks that can be activated concurrently within a batch under mapping $f$. \update{This term counts banks that contain useful operands for the current PIM batch. Since a pseudo-channel contains multiple bank groups and each bank group contains multiple banks, both bank-group bits and bank bits determine how many useful banks are populated by the mapped tensor data.} This number is determined by the address bits assigned to levels at or above the bank (e.g., bank group, channel):
\begin{equation}
    \#_\text{active\_banks}(f) = 2^{\sum_{i \in \mathcal{L}_{\ge\text{bank}}} \sum_{j=0}^n \mathbf{X}_{i,j}},
    \label{eq:level-mapping}
\end{equation}
with $\mathcal{L}_{\ge\text{bank}}$ denoting all such memory levels.

The second term, $T_{\text{ref}} / T_{\text{bank}}(f)$, reflects intra-bank access efficiency. Here, $T_{\text{ref}}$ is a reference latency representing ideal access conditions that only consider time to access required rows and columns in a bank: 
\begin{equation}
    T_{\text{ref}} = \frac{2^{p_n+p_k}}{\#_\text{active\_banks}(f) \cdot \mathbb{N}_{row}} \cdot t_{\text{RCD}} + \mathbb{N}_{col} \cdot t_{\text{CCD}}, 
\label{eq:pim-row-activation}
\end{equation}
where $\mathbb{N}_{row}$ and $\mathbb{N}_{col}$ denote the number of rows in each bank and the column numbers per row.

We model the actual latency $T_{\text{bank}}(f)$ by taking the maximum of (i) the intra-bank DRAM access time and (ii) the time to stream the external operand(s)/results through the channel:
\begin{equation}
    T_{\text{bank}}(f) = \max \left\{T_{access}^{bank}(f),T_{io}(f)\right\}
\end{equation}
\update{The max operator represents a steady-state pipelined bottleneck model for offline mapping search. DRAM-array access and external operand/result streaming use different datapaths and can be overlapped across PIM batches, so the slower stage bounds steady-state throughput. The final cycle-level evaluation still enforces the full HBM-PIM timing model.}
The first term, $T_{access}^{bank}(f) = 2^{\sum_j X_{\text{row},j}} \cdot t_{\text{RCD}} + 2^{\sum_j X_{\text{col},j}} \cdot t_{\text{CCD}}$, captures DRAM-internal latency, counting the number of distinct row and column activations per bank under $f$.
The second term models the time to supply input/output data for PIM computing via external channels can be expressed as 
\begin{equation}
    T_{io}(f)=\frac{2^{p_n}}{2^{\sum_{j\in [0,p_n-1]}\mathbf{X}_{ch,j}} \cdot B_{\text{ch}}} + \frac{2^{p_k}}{2^{\sum_{j\in [p_n,n]}\mathbf{X}_{ch,j}} \cdot B_{\text{ch}}}, 
\label{eq:io-latency}
\end{equation}
where $p_n$ and $p_k$ are input/output data dimension, $2^{\sum_{j\in [0,p_n-1]}\mathbf{X}_{ch,j}}$ and $2^{\sum_{j\in [p_n,n]}\mathbf{X}_{ch,j}}$ are the numbers of channels spanned by $p_n$ and $p_k$, respectively. Moreover, $B_{\text{ch}}$ denotes per-channel bandwidth.

\subsubsection{Mapping Optimization}
Based on the two objectives, we solve a per-superpage multi-objective mixed-integer linear programming problem: 
\begin{equation}
    \max_{f} \ \text{NPU}(f) + \alpha \cdot \text{PIM}(f),
\end{equation}
where $\alpha$ is chosen from operator characteristics through lightweight profiling that captures operator compute load variability (e.g., arithmetic intensity, reuse window). Using an open-source solver COIN-OR~\cite{COINOR_Projects} with PulP~\cite{pulp-solver} api, PFM materializes each solution as a Mapping Descriptor inside the 2MB page. \update{For the evaluated HBM configuration, the feasible mapping space is roughly $10^{15}$ candidates, and the COIN-OR/PuLP solver takes up to 30 minutes per tensor. This search is performed offline during profiling/setup and does not affect runtime inference.}

\subsection{Dual-view Addressing \& Scheduling}\label{sec:memory-aware scheduling}

To realize dual-optimized access without duplication, PFM augments the NPU-side memory controller with two lightweight components: (i) a Dual-View Address Remapping Unit (ARU) and (ii) a Flexible Access Scheduler (FAS). ARU converts physical addresses into device-specific hardware coordinates (NPU-view or PIM-view) using per-page descriptors; FAS then schedules the requests to maintain high utilization for both NPU and PIM.

\subsubsection{Address Remapping Unit (ARU)}
As illustrated in the left part of \fig~\ref{fig:scheduler}, ARU enables flexible address mapping through two key components: the Mapping Descriptor Table (MDT) and an address generation unit (AGU). The MDT stores descriptions of mapping functions, each specifying how physical address bits are assigned across memory levels. At runtime, the address generation unit translates incoming addresses into hardware addresses by applying the selected mapping rule from the MDT. The mapping is obtained during \texttt{PFMMalloc} with our dual-view mapping optimization.

To minimize modifications to existing system interfaces, memory requests retain their standard format, including command (cmd) and address (addr) fields. PFM extends these requests by embedding two additional fields within reserved bits: (1) a \texttt{MapID} field that identifies the desired address mapping scheme, and (2) a PIM operation type field specifying extended PIM instructions, such as matrix multiply-accumulate (\texttt{PIMMAC})~\cite{lee20221ynm,lee2021hardware}. For the PIM instructions, the exact bit allocation depends on the underlying PIM architecture. For instance, in a design similar to AiM~\cite{lee20221ynm} supporting 16 PIM instructions, unused bits in the request, e.g, CA[3:0], can indicate the PIM instruction type. Meanwhile, other reserved bits, e.g., CA[8:4], can encode the \texttt{MapID}. All other fields remain unmodified and regular DRAM requests set reserved bits to 0, ensuring compatibility with conventional memory protocols.

\textbf{Address generation pipeline.} 
Upon receiving a request, the ARU parses the command field to extract \texttt{MapID} and retrieves the corresponding mapping configuration from the MDT. The address generation unit then decomposes the address according to this mapping. For example, if addr[13:10] is designated as the channel index, it right-shifts the address by 10 bits and applies a mask to extract the 4-bit channel value. Physical addresses for other memory levels are computed similarly. Depending on the type of instruction, the ARU routes the request to one of two dedicated buffers, providing device-specific access for downstream scheduling.

\subsubsection{Flexible Access Scheduler (FAS)}
After completing address translation, PFM employs a dual-view scheduler to manage NPU and PIM requests separately while coordinating their execution via priority arbitration. As shown in \fig~\ref{fig:scheduler}, the scheduler includes request queues, NPU and PIM CMD selectors, a bank state tracker, and a priority arbiter feeding per-channel dispatch FIFOs.
\update{This runtime scheduler is what composes per-superpage mappings into a globally coordinated memory system. The mapping optimizer exposes efficient access candidates for each tensor page, while FAS observes shared channel/bank state and request queues to arbitrate among concurrent tensors, preventing independent local mappings from issuing uncoordinated requests to the same memory resources.}

For NPU requests, the scheduler applies an access reordering strategy guided by offline mapping. The dual-view mapping from \texttt{PFMMalloc} naturally exposes multi-channel parallelism, which the scheduler exploits by prioritizing instruction groups that span multiple channels to maximize bus-level concurrency. Within each channel, the bank state tracker identifies active rows, allowing row-hit accesses to be scheduled first and reducing row conflicts. These choices are made by lightweight comparators in the NPU CMD selector, with per-channel queues buffering pending requests. To mitigate fragmentation from non-contiguous accesses, a relayout buffer aggregates scattered segments across channels and reconstructs them into logical tensor order before delivery to the NPU, improving stream continuity.

For PIM requests, the scheduler leverages the bank state tracker to identify and aggregate instructions targeting different banks to reduce command overhead and enhance parallel efficiency. These are batched into per-channel groups and stored in the PIM CMD selector. Lightweight comparators within the selector quickly identify parallelizable candidates, enabling rapid dispatch for arbitration.

\textbf{Priority arbitration.}
The arbiter produces the joint schedule. It first services mandatory DRAM maintenance (e.g., refresh) to preserve timing guarantees. When both queues are non-empty, it prioritizes NPU memory accesses, as computation in NPU tends to be bursty and compute-intensive~\cite {Intel_NPU_decode_limited_by_compute, Dong2025_async_KV_prefetching}. After the NPU’s data bursts complete and it enters a compute-bound window, the arbiter opportunistically drains PIM batches as background work. Symmetrically, during prolonged PIM execution, the arbiter exploits idle banks to preemptively serve urgent NPU loads/stores. Selected commands are then staged into per-channel dispatch FIFOs and executed by the corresponding channel controllers.

  \section{LLM Inference with PFM}\label{sec:llm-inference-pfm}

\begin{figure}
    \centering
    \includegraphics[width=0.98\linewidth]{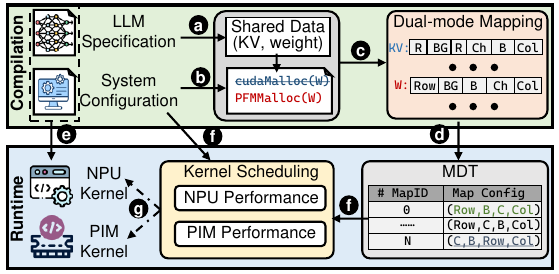}
    \caption{LLM inference workflow with PFM. }
    \label{fig:pfm-llm}
\vspace{-\baselineskip}
\end{figure}

Having introduced PFM’s mechanisms, we now detail how to instantiate PFM in an LLM inference framework. The end-to-end workflow is designed to adapt dynamically to shifting workload demands while avoiding costly data duplication. 

\begin{figure*}
    \centering
    \includegraphics[width=\linewidth]{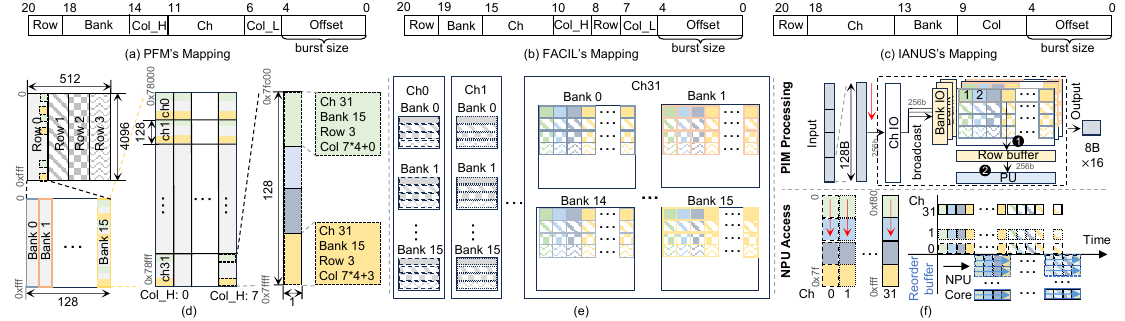}
    \caption{The case study of PFM design. \textbf{(a)} Illustration of the mapping strategy acquired by PFM's dual-view mapping for one tensor in Mixtral-8$\times$7B; Mapping strategies of \textbf{(b)} FACIL and \textbf{(c)} IANUS; \textbf{(d)} Illustration of weight tensor address with PFM's mapping; \textbf{(e)} Tensor layout in the memory; \textbf{(f)} Procedure of PIM computation \& Timeline for NPU data access.}
    \label{fig:case-study}
\vspace{-\baselineskip}
\end{figure*}

\subsection{Workflow}\label{sec:workflow}

As illustrated in \fig~\ref{fig:pfm-llm}, the workflow begins with a lightweight \textbf{offline} compilation phase that analyzes \circled{a} the LLM specification to identify tensors shared by both NPU and PIM, primarily KV caches and weights. Then it generates the needed mapping templates. Developers replace standard allocation calls (e.g., \texttt{cudaMalloc}) with \texttt{PFMMalloc} \circled{b} to trigger dual-view mapping \circled{c} and bind an optimized PA-to-HA mapping for each tensor. These mappings are registered in the MDT \circled{d} for runtime access. The compiler \circled{e} also generates both NPU and PIM kernels for the same LLM operators, enabling runtime flexible execution.

At \textbf{runtime}, PFM’s scheduler \circled{f} monitors the execution context and uses a performance model to dynamically assign each operator to the NPU or PIM. The selected kernel is launched \circled{g} with associated layout metadata, enabling the hardware to deliver NPU-optimized, channel-interleaved streams or PIM-oriented, bank-parallel broadcasts. This “allocate once, run anywhere” design seamlessly unifies data placement and device scheduling with minimal modifications to the existing inference framework. 

\subsection{NPU-PIM Kernel Scheduling}\label{sec:runtime-scheduling}

\subsubsection{Runtime Device Selection}\label{sec:runtime-selection}
To manage runtime variability during LLM inference, PFM employs a selection mechanism based on lightweight yet accurate performance modeling. At its core is a roofline model based predictor that estimates execution time for each operator on both NPU and PIM.

For illustration, consider a dynamically activated expert with weight matrix $W \in \mathbb{R}^{d_{\text{in}} \times d_{\text{out}}}$, processing $n$ tokens at the current step. The NPU execution time is modeled as:

\begin{equation}
T_{\text{NPU}} = \max\left( \frac{n \cdot d_{\text{in}} \cdot d_{\text{out}}}{\mathcal{P}_{\text{NPU}}},\ \frac{W}{\mathcal{B}_{\text{NPU}}} \right),
\end{equation}
where $\mathcal{P}_{\text{NPU}}$ and $\mathcal{B}_{\text{NPU}}$ denote peak compute throughput (FLOPs/s) and effective memory bandwidth (Bytes/s). This captures the NPU’s compute-memory trade-off: execution is compute-bound for large $n$, but limited by weight loading otherwise.

For PIM, performance depends on both compute and data layout quality:
\begin{equation}
T_{\text{PIM}} = \max\left( \frac{n \cdot d_{\text{in}} \cdot d_{\text{out}}}{\mathcal{P}_{\text{PIM}}^{\text{eff}}},\ \frac{W}{\mathcal{B}_{\text{PIM}}^{\text{eff}}} \right) + T_{\text{setup}},
\end{equation}
where effective throughput and bandwidth are scaled by the bank-level parallelism $\gamma \in [0,1]$, as $\mathcal{B}_{\text{PIM}}^{\text{eff}} = \gamma \cdot \mathcal{B}_{\text{PIM}}^{\text{peak}}$. When $\gamma \to 1$, PIM approaches peak bandwidth; when $\gamma \ll 1$, poor alignment leaves many banks idle. $T_{\text{setup}}$ accounts for fixed PIM command injection and bank-locking overhead, with both $\gamma$ and $T_{\text{setup}}$ determined offline. At runtime, the scheduler compares $T_{\text{NPU}}$ and $T_{\text{PIM}}$ and dispatches the operator to the device with lower predicted latency.

\subsubsection{Runtime Kernel Launch}
At runtime, PFM implements a unified heterogeneous execution engine that intercepts operator calls from the inference framework. Guided by the performance model, it dynamically selects the optimal device (either NPU or PIM) and invokes the computing kernel through a unified \texttt{PFMlaunch} interface. NPU execution proceeds through standard CUDA/HIP stream submission, whereas PIM execution serializes the kernel ID and parameters into a control message, delivered to the PIM array.
This entire process is transparent to upper layers, enabling developers to use standard MoE modules without modification or awareness of the underlying hardware heterogeneity.


  \section{Case Study}\label{sec:case-study}
We present a representative execution case from Mixtral-8$\times$7B~\cite{jiang2024mixtral} to illustrate how PFM’s dual-view mapping enables efficient data access for both NPU and PIM. We analyze a single expert weight tensor of size $4096 \times 14336$ (INT8, 56\,MB), stored in column-major order. The tensor is divided into 28 superpages, each spanning $4096 \times 512$ elements. Our analysis focuses on the layout and access behavior of one such page.

\subsection{Optimized Tensor Layout}

\subsubsection{Addressing \& Layout with PFM}
To simplify the PA-to-HA mapping, we merge the bank and bank group levels and treat pseudo-channels as channels. The resulting HBM configuration comprises 32 channels, each with 16 banks, with each row containing 1 KB and a 32-byte burst. Under this configuration, the dual-view mapping is shown in \fig~\ref{fig:case-study}(a), with the corresponding addressing in \fig~\ref{fig:case-study}(d). Each 2 MB page is partitioned into four $4096 \times 128$ sub-tensors, each mapped to a fixed row index. Within a sub-tensor, data are striped across banks such that the same bank index across channels stores a $4096 \times 8$ tile, while the column dimension is further interleaved to distribute each $4096 \times 1$ vector evenly across 32 channels. The resulting physical layout is illustrated in \fig~\ref{fig:case-study}(e).


\subsubsection{Benefits of PFM Mapping}
PFM's address mapping significantly enhances both NPU access and PIM computing efficiency. By assigning channel (ch) bits to the lower address range [11:7], PFM enables all-channel parallelism whenever more than 2048B are accessed simultaneously (\eqn~\eqref{eq:npu-parallel}). The row bits are the top two positions [20:19], allowing the entire 2MB page to be accessed with only four times of row activation per bank, thus minimizing row activation overhead for both NPU (\eqn~\eqref{eq:bank-time}-\eqref{eq:row-activation}) and PIM operations (\eqn~\eqref{eq:pim-row-activation}).
Moreover, by partitioning the column field into Col\_H and Col\_L and embedding them at both ends of the channel field, PFM reduces I/O transfer overhead during PIM computation (\eqn~\eqref{eq:io-latency}). Col\_L enables concurrent data transfers across channels, whereas only part of the column field is assigned to Col\_H to minimize cross-channel reductions during GEMV execution, further improving processing efficiency. 


\subsection{Efficient Tensor Access}
\fig~\ref{fig:case-study}(f) illustrates tensor processing on both devices.
On the PIM side, computation runs in all-bank mode: each channel’s I/O path supplies the GEMV input vector, and banks perform partial inner reductions. The global reduction on the base die~\cite{park2024attacc} accounts for only a small fraction (1/128) of total execution overhead and thus does not stall the pipeline.
On the NPU side, 32 channels each with 32-bit width operate in parallel to saturate the 1024-pin interface \cite{a100paper}, and the relayout buffer in the PFM controller reassembles data streams before passing them to the NPU core. With the PFM mapping, each 128 B segment lies within a single channel and requires four bursts for full retrieval, adding three extra bursts (24 cycles or $\sim$15 ns at 3.2 Gbps, double rate). With multiple channels operating in parallel and operator latency typically exceeding 100 µs~\cite{zhang2024llmcompass}, this delay is negligible and does not affect throughput.
  \section{Evaluation}

\subsection{Experimental Setup}\label{sec:experimental-setup}


\noindent \textbf{Baselines.}
We evaluate PFM against three representative systems: NPU-only, PSM and PUM. 

\begin{itemize}[leftmargin=*, nosep]
    \item \textbf{NPU-only} is a GPU system based on vLLM~\cite{kwon2023efficient}, using NVIDIA A100-80G GPUs for LLM inference~\cite{a100paper}. It reflects current industry practice and serves as our primary performance reference.

    \item \textbf{PSM} \update{employs separated memory management, following a NeuPIMs-like~\cite{heo2024neupims} setup. Here, ``NeuPIMs-like'' means that the memory space is divided into NPU-side and PIM-side regions: model weights are placed in NPU regions, while KV cache is placed in PIM regions. This static allocation facilitates PIM acceleration for memory-bound decode-time attention but limits system flexibility.}

    \item \textbf{PUM} \update{employs unified memory with PIM-friendly address mapping for all tensors, following the PIM-side mapping used in FACIL~\cite{seo2025facil}. This baseline exposes shared data to PIM without separating the NPU and PIM memory regions, and it isolates the performance impact of unified memory with a PIM-biased layout. In small-batch settings, PUM is consistent with the full FACIL mechanism because FACIL also stores all tensors in a PIM-friendly layout to favor PIM execution. In large-batch settings, FACIL maps FFN layers to NPU-friendly layout and KV cache to PIM-friendly layout, which becomes equivalent to PSM in our evaluation. Therefore, large-batch FACIL performance can be inferred from the PSM baseline, while PUM reports the cost of an all-PIM-friendly unified-memory design.}

\end{itemize}

\begin{table}[t]
    \centering
    \caption{Configuration Details of HBM-PIM.}
    \resizebox{\linewidth}{!}{
    \begin{tabular}{c|c|c}
    \hline
    \multicolumn{3}{c}{\textbf{HBM Parameters}} \\
    \hline
     \multicolumn{3}{c}{HBM2e\_3.2Gbps, 16GB/HBM, 8 channel/HBM}\\
     \multicolumn{3}{c}{2 pseudo channel/ch, 2 rank/pch, 4 bank groups/rank, 4 bank/BG}\\
     \hline
    \multicolumn{3}{c}{\textbf{HBM Timing (ns)}} \\
    \hline
        \multicolumn{3}{c}{tRP = 15, tRCD = 14, tRAS = 33, tRRD\_L = 6, tWR = 16}\\
     \multicolumn{3}{c}{tCCD\_S = 1.25, tCCD\_L = 3.50, tREFI = 3900, tRFC = 350, tFAW = 30}\\
    \hline
    \multicolumn{3}{c}{\textbf{Processing Unit Parameters}} \\
    \hline
     \multicolumn{3}{c}{PU: 666 MHz; 1 PU per bank (1024 PU per stack); 12.8TFLOPS per stack}\\
     \hline
    \end{tabular}
    }
    \label{tab:pim-config}
\vspace{-\baselineskip}
\end{table}
\begin{figure*}[t]
    \centering
    \includegraphics[width=0.95\linewidth]{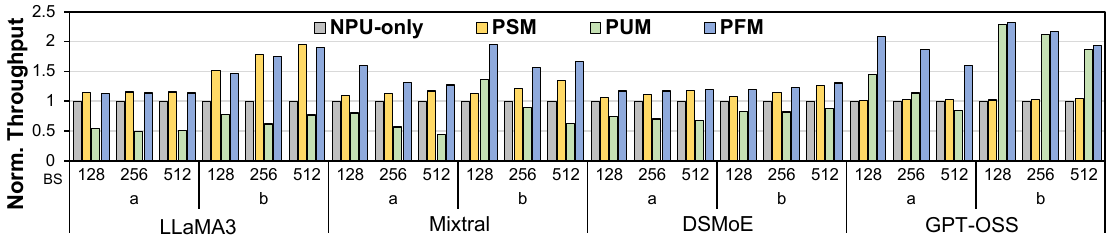}
    \caption{\update{Normalized end-to-end inference throughput (tokens/s) on LLaMA3, Mixtral, DSMoE and GPT-OSS. a means $L_{in}=1024$, $L_{out}=128$; b means $L_{in}=128$, $L_{out}=1024$.}}
    \label{fig:throughput}
\end{figure*}

\noindent
For fair comparison, PFM uses the same hardware as the NPU-PIM baselines: all HBM modules are replaced with HBM-PIM stacks. All systems share identical HBM capacity and A100 computational resources, differing only in memory management strategy.
\update{For all quantitative motivation and evaluation figures, NPU-PIM systems use A100-class compute resources with HBM replaced by AttAcc-style HBM-PIM stacks. The main evaluation baselines follow the PSM/PUM definitions above, while the additional motivation-only strategies are defined in Section~\ref{sec:motivation}.}

\noindent \textbf{Simulation.}
\update{We developed a system-level simulator integrating GPGPU-Sim 4.0~\cite{khairy2020accel, bakhoda2009analyzing} and Ramulator 2.0~\cite{luo2023ramulator, ramulator2.0}. Since GPUs are among the mainstream hardware platforms for production LLM inference, we evaluate the NPU side using GPUs. We run the evaluated models with vLLM, collect the CUDA kernels issued during execution, and simulate the compute-related kernels with GPGPU-Sim. During simulation, we also record their memory access patterns. The resulting traces are replayed through Ramulator 2.0 to model NPU-side memory timing under different PA-to-HA mappings.}

\update{For operators scheduled to PIM, we generate PIM memory requests from operator shapes, tensor layouts, and selected mappings, following the trace-generation methodology of AttAcc~\cite{park2024attacc}. We enhance Ramulator 2.0 with the HBM-PIM timing model, PIM execution constraints, and PFM's ARU/FAS controller logic. The modeled PIM constraints include SIMD width and register-level reuse. The lookup/address-remapping latency measured from our RTL implementation is added to the affected memory-request latency, so the reported performance includes MDT lookup, address generation, and scheduler-side handling overheads.}

\update{End-to-end latency and throughput are composed from the simulated prefill, decode, attention, FFN, and MoE expert kernels, using sampled expert-activation behavior for MoE models. We validate the simulator against real A100 measurements~\cite{dgx-a100} and open-source NPU-PIM simulators~\cite{heo2024neupims, park2024attacc}; across the validation cases, the simulation error ranges from -15.57\% to 0\%, with an average error of -5.14\%.}

\noindent \textbf{Hardware Specification.} 
We adopt an HBM-PIM architecture, with detailed parameters summarized in \tab \ref{tab:pim-config}. Each memory stack operates at 3.2 Gbps, providing 16 GB capacity and 8 channels per stack, with timing parameters following vendor specifications~\cite{JEDEC_JESD235D, Samsung_HBM2e_3.2Gbps}. The PIM design follows the near-bank architecture used in AttAcc~\cite{park2024attacc}, where each bank is coupled with a Processing Unit (PU) running at 666 MHz, delivering a peak throughput of 12.8 TFLOPS per stack.

\begin{table}
\centering
\caption{Evaluated LLM configurations.}
\label{tab:llm-configuration}
\resizebox{\linewidth}{!}{
\begin{tabular}{|c|c|c|c|c|}
\hline
\textbf{Model} & \textbf{\# Layers} & \textbf{\# Dim} & Experts & \textbf{\# Hardware Config.} \\ \hline
LLaMA3-8B~\cite{dubey2024llama3}  & 32  & 4096 & Dense & 1$\times$ NPU \\ \hline
DeepSeekMoE-16B~\cite{dai2024deepseekmoe} & 28 & 2048 & 2+64 (2+4) & 1$\times$ NPU \\ \hline
Mixtral 8$\times$7B~\cite{jiang2024mixtral}  & 32 & 4096 & 8 (2)  & 2$\times$ NPUs  \\ \hline
GPT-OSS-120B~\cite{gpt-oss}  & 36 & 4096 & 128 (4)  & 4$\times$ NPUs\\ \hline
\end{tabular}
}
\vspace{-\baselineskip}
\end{table}

\noindent \textbf{Models.}
We use four representative LLMs for evaluation: LLaMA3-8B (LLaMA3)~\cite{dubey2024llama3}, DeepSeekMoE-16B (DSMoE)~\cite{dai2024deepseekmoe}, Mixtral 8$\times$7B (Mixtral)~\cite{jiang2024mixtral}, and GPT-OSS-120B (GPT-OSS)~\cite{gpt-oss}.
LLaMA3-8B employs a dense architecture, whereas the remaining models are based on the MoE paradigm, dynamically activating a subset of experts. Specifically, DSMoE activates 2 shared experts and 4 out of 64 (4/64) routed experts per token; Mixtral activates 2/8 experts; and GPT-OSS selects 4/128 experts. The details of each model, along with the corresponding number of NPUs, are summarized in Table~\ref{tab:llm-configuration}. Each NPU provides 80 GB memory capacity. The diverse models validate PFM’s adaptability across workload profiles.

\noindent \textbf{Workloads.}
Our evaluation focuses on inference latency and throughput under diverse scenarios. Batch sizes vary to capture both latency-sensitive services (with small batch sizes) and throughput-oriented processing (with large batch sizes) scenarios. Furthermore, we evaluate performance across a range of input and output sequence lengths ($L_{in}$ and $L_{out}$) to capture the impact of varying demands. 


\subsection{End-to-end Performance}\label{sec:performance}

\begin{figure}
    \centering
    \includegraphics[width=\linewidth]{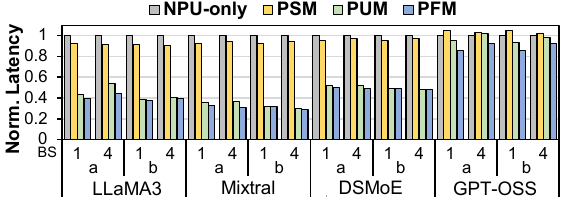}
    \caption{\update{The normalized end-to-end inference latency with small batch sizes (BS=1 and 4). a means $L_{in}=1024$, $L_{out}=128$; b means $L_{in}=128$, $L_{out}=1024$.}}
    \label{fig:latency}
\vspace{-\baselineskip}
\end{figure}

\noindent \textbf{Latency-sensitive Performance.}
\fig~\ref{fig:latency} reports end-to-end inference latency under small batch sizes (1 and 4), representing latency-critical scenarios. Compared to NPU-only, PSM, PUM, and PFM achieve average speedups of 1.04$\times$, 2.10$\times$, and 2.23$\times$, respectively. 
PSM provides modest gains by accelerating decode-time attention on PIM, which becomes more pronounced for long output sequences ($L_{\text{out}}=1024$). PUM further improves performance by offloading the entire decode phase to PIM, which is well-suited for memory-intensive operations in small-batch settings. However, its PIM-biased memory layout degrades NPU bandwidth during the prefill phase. As a result, when prefill dominates execution (e.g., $L_{\text{in}}=1024$, $L_{\text{out}}=128$), PUM shows limited benefit, and PFM achieves up to $1.21\times$ speedup over PUM.
Across all configurations, PFM consistently outperforms both baselines. By extending PIM acceleration to additional decode operators and adopting a dual-optimized layout that preserves NPU efficiency during prefill, PFM avoids the performance trade-offs observed in PUM, delivering robust latency improvements across diverse input–output regimes.

\noindent \textbf{Throughput-oriented Performance.}
\fig~\ref{fig:throughput} reports inference throughput under large batch sizes (128–512) for both dense and MoE models. Compared to NPU-only, PSM, and PUM, PFM improves average throughput by 1.20$\times$, 1.06$\times$, and 1.55$\times$ (up to 1.95$\times$, 2.29$\times$, and 2.32$\times$), respectively.
These gains primarily stem from PFM’s runtime and performance-aware scheduling. Under large batches, while linear layers benefit from increased data reuse~\cite{park2024attacc, heo2024neupims}, attention remains memory-bound due to batch-independent KV cache accesses and thus continues to benefit from PIM acceleration. PFM further offloads sparsely activated MoE experts to PIM, where limited invocation frequency leads to low arithmetic intensity. This improves resource utilization and reduces tail latency, thereby boosting overall throughput. Importantly, PFM maintains high NPU efficiency for compute-intensive operators through its dual-optimized memory layout.
In contrast, PSM accelerates only attention and thus provides limited benefits for MoE models. PUM suffers from degraded NPU performance due to its biased memory layout, making it less effective for dense models and highly NPU-dominated workloads. By enabling flexible operator-to-device mapping without sacrificing NPU efficiency, PFM consistently delivers higher and more scalable throughput across diverse scenarios.

\subsection{Performance Analysis}\label{sec:analysis}
To gain deeper insight into the performance gains of the PFM-based LLM inference system, we conduct a comprehensive analysis of its behavior across diverse workloads.

\begin{figure}
    \centering
    \includegraphics[width=0.98\linewidth]{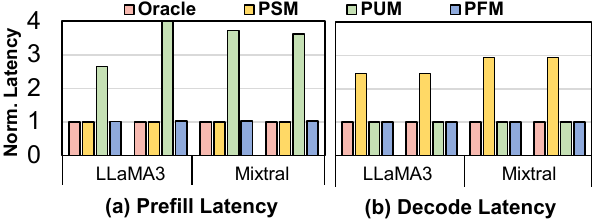}
    \caption{\update{Normalized latency on prefill and decode phases with a single batch. Prefill uses $L_{in}=1024$, $L_{out}=128$, and decode uses $L_{in}=128$, $L_{out}=1024$. Oracle denotes an upper bound with ideal operator placement and layout support on the same NPU-PIM hardware.}}
    \label{fig:prefill-decode}
\vspace{-\baselineskip}
\end{figure}

\noindent \textbf{Prefill \& Decode.}
\fig~\ref{fig:prefill-decode} compares prefill and decode latency under single-batch workloads. The oracle assumes ideal memory access without layout-induced overhead, serving as the theoretical upper bound. As shown in \fig~\ref{fig:prefill-decode}(a), PFM reduces prefill latency by $3.43\times$ over PUM through its dual-optimized memory layout, which maintains high bandwidth for NPU-bound operations. However, since decoding is autoregressive and often dominates end-to-end latency, this advantage translates to a smaller overall speedup, consistent with the end-to-end results in \fig~\ref{fig:throughput}. During decoding in \fig~\ref{fig:prefill-decode}(b), PFM outperforms PSM by supporting broader operator offloading. Since decode-phase operations exhibit low arithmetic intensity (1 FLOP/byte), they are well-suited for PIM execution. Against the oracle design, PFM achieves 96.8\% of the end-to-end efficiency. This near-optimal performance demonstrates that its dual-optimized memory layout effectively reconciles the conflicting access patterns of NPU and PIM.

\begin{figure}
    \centering
    \includegraphics[width=0.98\linewidth]{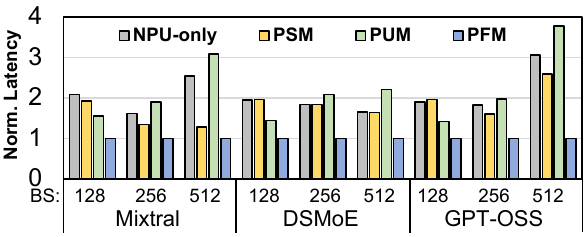}
    \caption{The normalized inference latency on MoE blocks.}
    \label{fig:moe}
\end{figure}

\noindent \textbf{Performance on MoE.}
\fig~\ref{fig:moe} evaluates PFM on MoE models. Instead of offloading experts entirely on NPUs, PSM adopts a typical static allocation strategy~\cite{zhong2025hybrimoe, cao2025moe}, which decides the computing devices with offline profiling. Compared to NPU-only, PSM, and PUM, PFM reduces MoE block latency by $1.76\times$, $2.01\times$, and $2.05\times$, respectively. The significant activation variation of different experts leads to \update{low efficiency} in NPU-only and PUM designs. PSM with static allocation, while attempting to allocate each expert to its suitable device, cannot adapt to step-to-step variations\footnote{We provide real MoE expert activation traces at  (\url{https://anonymous.4open.science/r/PFM-9F2D}).} in activation patterns, resulting in persistent imbalances. In contrast, PFM dynamically schedules experts across NPU and PIM based on runtime behavior, mitigating load imbalance and improving hardware utilization. This adaptive execution enables consistently higher and more scalable performance for MoE-based LLM inference.

\subsection{Bandwidth Utilization}\label{sec:utilization}


We further evaluate the bandwidth utilization of data access on both the NPU and PIM devices, covering QKV weight, expert weight, and KV cache access. As shown in Table \ref{tab:utilization}, we compare the bandwidth utilization of NPU and PIM when running Mixtral. Unlike PUM, which achieves high access bandwidth (an average 95.33\%) only at the PIM side, PFM effectively balances bandwidth efficiency for both NPU-side and PIM-side accesses. Overall, PFM achieves a bandwidth utilization of 92.10\% on NPU accesses and 94.85\% on PIM accesses. Similar trends are consistently observed across models, demonstrating the effectiveness of the proposed technique in enabling efficient large-scale LLM inference.

\begin{table}
\caption{Bandwidth utilization with diverse tensor shape during Mixtral inference. }
\centering
\resizebox{\linewidth}{!}{
\begin{tabular}{lccccc}
\hline
 & & \textbf{4K$\times$4K (QKV)} & \textbf{4K$\times$512 (KV cache)} & \textbf{4K$\times$14K (MoE)} & \textbf{14K$\times$4K (MoE)} \\ 
\hline
\multirow{2}{*}{PUM}& NPU & 14.33 \% & 23.52\% & 16.61\% & 15.26\% \\
& PIM & 95.12\% & 94.18\% & 96.29\% & 95.71\% \\
\hline
\multirow{2}{*}{PFM}& NPU & 93.94\% & 90.86\% & 93.64\% & 89.97\% \\
& PIM & 94.09\% & 93.53\% & 95.31\% & 96.48\% \\
\hline
\end{tabular}
}
\label{tab:utilization}
\end{table}

\subsection{Area Overhead}\label{sec:area}
We implemented and synthesized the ARU and FAS modules in the memory controller to evaluate area overhead, using RTL design and Synopsys Design Compiler~\cite{synopsys.org} with a 7nm predictive process kit~\cite{clark2016asap7}. The main contributors to the area are buffers for instruction/data and control logic. In the ARU, the MDT contains 32 registers. In the FAS, large request buffers (1 MB each for NPU and PIM) are provisioned to enable scheduling over a wider instruction window, each holding up to 128K memory requests. The NPU-PIM command selectors store 128 schedulable instructions per channel (1 KB buffer per channel), and the dispatch FIFO adopts the same configuration. A 64 KB relayout buffer supports multiple burst transfers for data reorganization. \update{The total area is 2.82 mm$^2$. Because ARU/FAS resides in the memory-controller path, the most relevant public estimate is the memory-controller/HBM-interface region. Based on die-photo floorplan estimates~\cite{handy2021future_low_latency_memory}, each A100 HBM2E interface occupies 11.4 mm$^2$. Since A100 uses five active HBM2E interfaces, the active memory-interface area is 57.0 mm$^2$, and ARU/FAS adds 4.95\% over this region. As a secondary scale reference, this corresponds to 0.34\% of the full A100 die~\cite{a100paper}.}

  \section{Discussion}
\noindent
\textbf{Compatibility.} PFM guarantees the consistency expected by existing NPU-PIM systems. For virtual memory management, PFM leverages the native 2 MB superpages already supported by modern GPUs~\cite{a100paper, h100paper}. In consequence, our design does not introduce modifications to the fundamental virtual-to-physical address translation. From the software stack, PFM is fully compatible with existing inference frameworks~\cite{sglang-project, vllm-project, paszke2019pytorch}, requiring only negligible modification on memory allocation. It allows PFM to deliver performance gains without disrupting the established software stack.

\noindent \textbf{Scalability.} 
PFM's dual-view mapping optimization and runtime scheduling are memory technology agnostic and can be applied to various memory technologies, including HBM, GDDR, and LPDDR. The approach is also largely independent of NPU architectures, making it compatible with a wide range of NPU designs~\cite{norrie2021design, liao2021ascend, liao2019davinci}. This flexibility enables broad applicability while preserving efficiency.

\noindent \textbf{Limitations}
\update{PFM relies on offline profiling and a finite number of MDT entries to store mapping templates. This is effective for common LLM serving deployments because tensors with similar shapes/layouts can share one MapID, and our 32-entry MDT is sufficient for multiple evaluated concurrent models. However, if expert activation patterns or the set of served models changes substantially, PFM may require re-profiling or a larger MDT to maintain mapping coverage.}

\section{Conclusion}

We present PFM, a unified and flexible memory management on NPU-PIM systems that addresses the significant data sharing and dynamic demands of LLM inference. Unlike prior designs that assume biased data access, PFM enables flexible, dual-view memory access by decoupling logical views from physical layouts via a software-hardware co-design. This design allows both NPU and PIM to efficiently access shared data without duplication. Implemented with minimal modifications to the memory controller and software stack, PFM improves up to $2.26 \times$ end-to-end LLM inference throughput improvement over the existing NPU-PIM system, demonstrating its necessity for efficient LLM serving.

\section*{Acknowledgment}
We sincerely thank the anonymous reviewers for their suggestions to improve the paper. This work was supported by the National Key R\&D Program of China: 2023YFB4404400.

\bibliographystyle{IEEEtranS}
\bibliography{ref}

@misc{sglang-project,
    title = {SGLang: Efficient Execution of Structured Language Model Programs},
    url = {https://github.com/sgl-project/sglang},
    author = {SGLang Team},
    month = {September},
    year = {2024}
}

@inproceedings{heo2024neupims,
  title={Neupims: Npu-pim heterogeneous acceleration for batched llm inferencing},
  author={Heo, Guseul and Lee, Sangyeop and Cho, Jaehong and Choi, Hyunmin and Lee, Sanghyeon and Ham, Hyungkyu and Kim, Gwangsun and Mahajan, Divya and Park, Jongse},
  booktitle={Proceedings of the 29th ACM International Conference on Architectural Support for Programming Languages and Operating Systems, Volume 3},
  pages={722--737},
  year={2024}
}

@inproceedings{park2024attacc,
  title={AttAcc! Unleashing the Power of PIM for Batched Transformer-based Generative Model Inference},
  author={Park, Jaehyun and Choi, Jaewan and Kyung, Kwanhee and Kim, Michael Jaemin and Kwon, Yongsuk and Kim, Nam Sung and Ahn, Jung Ho},
  booktitle={Proceedings of the 29th ACM International Conference on Architectural Support for Programming Languages and Operating Systems, Volume 2},
  pages={103--119},
  year={2024}
}

@inproceedings{seo2024ianus,
  title={IANUS: Integrated Accelerator based on NPU-PIM Unified Memory System},
  author={Seo, Minseok and Nguyen, Xuan Truong and Hwang, Seok Joong and Kwon, Yongkee and Kim, Guhyun and Park, Chanwook and Kim, Ilkon and Park, Jaehan and Kim, Jeongbin and Shin, Woojae and others},
  booktitle={Proceedings of the 29th ACM International Conference on Architectural Support for Programming Languages and Operating Systems, Volume 3},
  pages={545--560},
  year={2024}
}

@inproceedings{lee20221ynm,
  title={A 1ynm 1.25 V 8Gb, 16Gb/s/pin GDDR6-based accelerator-in-memory supporting 1TFLOPS MAC operation and various activation functions for deep-learning applications},
  author={Lee, Seongju and Kim, Kyuyoung and Oh, Sanghoon and Park, Joonhong and Hong, Gimoon and Ka, Dongyoon and Hwang, Kyudong and Park, Jeongje and Kang, Kyeongpil and Kim, Jungyeon and others},
  booktitle={2022 IEEE International Solid-State Circuits Conference (ISSCC)},
  volume={65},
  pages={1--3},
  year={2022},
  organization={IEEE}
}

@article{roziere2023code,
  title={Code llama: Open foundation models for code},
  author={Roziere, Baptiste and Gehring, Jonas and Gloeckle, Fabian and Sootla, Sten and Gat, Itai and Tan, Xiaoqing Ellen and Adi, Yossi and Liu, Jingyu and Sauvestre, Romain and Remez, Tal and others},
  journal={arXiv preprint arXiv:2308.12950},
  year={2023}
}

@article{chiang2023vicuna,
  title={Vicuna: An open-source chatbot impressing gpt-4 with 90\%* chatgpt quality},
  author={Chiang, Wei-Lin and Li, Zhuohan and Lin, Zi and Sheng, Ying and Wu, Zhanghao and Zhang, Hao and Zheng, Lianmin and Zhuang, Siyuan and Zhuang, Yonghao and Gonzalez, Joseph E and others},
  journal={See https://vicuna. lmsys. org (accessed 14 April 2023)},
  volume={2},
  number={3},
  pages={6},
  year={2023}
}

@inproceedings{xie2024waitgpt,
  title={WaitGPT: Monitoring and Steering Conversational LLM Agent in Data Analysis with On-the-Fly Code Visualization},
  author={Xie, Liwenhan and Zheng, Chengbo and Xia, Haijun and Qu, Huamin and Zhu-Tian, Chen},
  booktitle={Proceedings of the 37th Annual ACM Symposium on User Interface Software and Technology},
  pages={1--14},
  year={2024}
}

@article{kabakucs2024battle,
  title={The battle of Chatbot Giants: an experimental comparison of ChatGPT and Bard},
  author={Kabaku{\c{s}}, Abdullah Talha and Dogru, {\.I}brahim},
  journal={International Journal of Engineering Research and Development},
  volume={16},
  number={2},
  pages={679--691},
  year={2024},
  publisher={Kirikkale University}
}

@article{dubey2024llama3,
  title={The llama 3 herd of models},
  author={Dubey, Abhimanyu and Jauhri, Abhinav and Pandey, Abhinav and Kadian, Abhishek and Al-Dahle, Ahmad and Letman, Aiesha and Mathur, Akhil and Schelten, Alan and Yang, Amy and Fan, Angela and others},
  journal={arXiv preprint arXiv:2407.21783},
  year={2024}
}

@misc{chat-gpt,
    title = {ChatGPT},
    url = {https://chatgpt.com/blog/chatgpt},
    author = {OpenAI},
    month = {September},
    year = {2023}
}

@misc{claude,
    title = {Introducing the next generation of claude.},
    url = {https://www.anthropic.com/news/claude-3-family},
    author = {The Claude Team},
    month = {September},
    year = {2024}
}

@inproceedings{he2025papi,
  title={Papi: Exploiting dynamic parallelism in large language model decoding with a processing-in-memory-enabled computing system},
  author={He, Yintao and Mao, Haiyu and Giannoula, Christina and Sadrosadati, Mohammad and G{\'o}mez-Luna, Juan and Li, Huawei and Li, Xiaowei and Wang, Ying and Mutlu, Onur},
  booktitle={Proceedings of the 30th ACM International Conference on Architectural Support for Programming Languages and Operating Systems, Volume 2},
  pages={766--782},
  year={2025}
}

@article{zhou2022mixture,
  title={Mixture-of-experts with expert choice routing},
  author={Zhou, Yanqi and Lei, Tao and Liu, Hanxiao and Du, Nan and Huang, Yanping and Zhao, Vincent and Dai, Andrew M and Le, Quoc V and Laudon, James and others},
  journal={Advances in Neural Information Processing Systems},
  volume={35},
  pages={7103--7114},
  year={2022}
}

@article{fedus2022switch,
  title={Switch transformers: Scaling to trillion parameter models with simple and efficient sparsity},
  author={Fedus, William and Zoph, Barret and Shazeer, Noam},
  journal={Journal of Machine Learning Research},
  volume={23},
  number={120},
  pages={1--39},
  year={2022}
}

@inproceedings{zhang2024llmcompass,
  title={Llmcompass: Enabling efficient hardware design for large language model inference},
  author={Zhang, Hengrui and Ning, August and Prabhakar, Rohan Baskar and Wentzlaff, David},
  booktitle={2024 ACM/IEEE 51st Annual International Symposium on Computer Architecture (ISCA)},
  pages={1080--1096},
  year={2024},
  organization={IEEE}
}

@misc{COINOR_Projects,
  title = {Projects – {COIN-OR}: Computational Infrastructure for Operations Research},
  author = {{COIN-OR Foundation}},
  howpublished = {\url{https://www.coin-or.org/projects/}},
  year = {2025}
}

@inproceedings{kwon2023efficient,
  title={Efficient memory management for large language model serving with pagedattention},
  author={Kwon, Woosuk and Li, Zhuohan and Zhuang, Siyuan and Sheng, Ying and Zheng, Lianmin and Yu, Cody Hao and Gonzalez, Joseph and Zhang, Hao and Stoica, Ion},
  booktitle={Proceedings of the 29th symposium on operating systems principles},
  pages={611--626},
  year={2023}
}

@inproceedings{hu2025lightllm,
  title={Lightllm: A versatile large language model for predictive light sensing},
  author={Hu, Jiawei and Jia, Hong and Hassan, Mahbub and Yao, Lina and Kusy, Brano and Hu, Wen},
  booktitle={Proceedings of the 23rd ACM Conference on Embedded Networked Sensor Systems},
  pages={158--171},
  year={2025}
}

@inproceedings{zhong2024distserve,
  title={$\{$DistServe$\}$: Disaggregating Prefill and Decoding for Goodput-optimized Large Language Model Serving},
  author={Zhong, Yinmin and Liu, Shengyu and Chen, Junda and Hu, Jianbo and Zhu, Yibo and Liu, Xuanzhe and Jin, Xin and Zhang, Hao},
  booktitle={18th USENIX Symposium on Operating Systems Design and Implementation (OSDI 24)},
  pages={193--210},
  year={2024}
}

@inproceedings{yun2024duplex,
  title={Duplex: A Device for Large Language Models with Mixture of Experts, Grouped Query Attention, and Continuous Batching},
  author={Yun, Sungmin and Kyung, Kwanhee and Cho, Juhwan and Choi, Jaewan and Kim, Jongmin and Kim, Byeongho and Lee, Sukhan and Sohn, Kyomin and Ahn, Jung Ho},
  booktitle={2024 57th IEEE/ACM International Symposium on Microarchitecture (MICRO)},
  pages={1429--1443},
  year={2024},
  organization={IEEE}
}

@inproceedings{kim2024sk,
  title={SK Hynix AI-Specific Computing Memory Solution: From AiM Device to Heterogeneous AiMX-xPU System for Comprehensive LLM Inference},
  author={Kim, Guhyun and Kim, Jinkwon and Kim, Nahsung and Shin, Woojae and Won, Jongsoon and Joo, Hyunha and Choi, Haerang and An, Byeongju and Shin, Gyeongcheol and Yun, Dayeon and others},
  booktitle={2024 IEEE Hot Chips 36 Symposium (HCS)},
  pages={1--26},
  year={2024},
  organization={IEEE Computer Society}
}

@inproceedings{li2025h2,
  title={H2-LLM: Hardware-Dataflow Co-Exploration for Heterogeneous Hybrid-Bonding-based Low-Batch LLM Inference},
  author={Li, Cong and Yin, Yihan and Wu, Xintong and Zhu, Jingchen and Gao, Zhutianya and Niu, Dimin and Wu, Qiang and Si, Xin and Xie, Yuan and Zhang, Chen and others},
  booktitle={Proceedings of the 52nd Annual International Symposium on Computer Architecture},
  pages={194--210},
  year={2025}
}

@article{wu2024pim,
  title={Pim gpt a hybrid process in memory accelerator for autoregressive transformers},
  author={Wu, Yuting and Wang, Ziyu and Lu, Wei D},
  journal={npj Unconventional Computing},
  volume={1},
  number={1},
  pages={4},
  year={2024},
  publisher={Nature Publishing Group UK London}
}

@article{chowdhery2023palm,
  title={Palm: Scaling language modeling with pathways},
  author={Chowdhery, Aakanksha and Narang, Sharan and Devlin, Jacob and Bosma, Maarten and Mishra, Gaurav and Roberts, Adam and Barham, Paul and Chung, Hyung Won and Sutton, Charles and Gehrmann, Sebastian and others},
  journal={Journal of Machine Learning Research},
  volume={24},
  number={240},
  pages={1--113},
  year={2023}
}

@article{workshop2022bloom,
  title={Bloom: A 176b-parameter open-access multilingual language model},
  author={Workshop, BigScience and Scao, Teven Le and Fan, Angela and Akiki, Christopher and Pavlick, Ellie and Ili{\'c}, Suzana and Hesslow, Daniel and Castagn{\'e}, Roman and Luccioni, Alexandra Sasha and Yvon, Fran{\c{c}}ois and others},
  journal={arXiv preprint arXiv:2211.05100},
  year={2022}
}

@inproceedings{lee2021hardware,
  title={Hardware architecture and software stack for PIM based on commercial DRAM technology: Industrial product},
  author={Lee, Sukhan and Kang, Shin-haeng and Lee, Jaehoon and Kim, Hyeonsu and Lee, Eojin and Seo, Seungwoo and Yoon, Hosang and Lee, Seungwon and Lim, Kyounghwan and Shin, Hyunsung and others},
  booktitle={2021 ACM/IEEE 48th Annual International Symposium on Computer Architecture (ISCA)},
  pages={43--56},
  year={2021},
  organization={IEEE}
}

@article{luo2023ramulator,
  title={Ramulator 2.0: A modern, modular, and extensible dram simulator},
  author={Luo, Haocong and Tu{\u{g}}rul, Yahya Can and Bostanc{\i}, F Nisa and Olgun, Ataberk and Ya{\u{g}}l{\i}k{\c{c}}{\i}, A Giray and Mutlu, Onur},
  journal={IEEE Computer Architecture Letters},
  volume={23},
  number={1},
  pages={112--116},
  year={2023},
  publisher={IEEE}
}

@inproceedings{zhao2024pim,
  title={Um-pim: Dram-based pim with uniform \& shared memory space},
  author={Zhao, Yilong and Gao, Mingyu and Liu, Fangxin and Hu, Yiwei and Wang, Zongwu and Lin, Han and Li, Ji and Xian, He and Dong, Hanlin and Yang, Tao and others},
  booktitle={2024 ACM/IEEE 51st Annual International Symposium on Computer Architecture (ISCA)},
  pages={644--659},
  year={2024},
  organization={IEEE}
}

@inproceedings{seo2025facil,
  title={FACIL: Flexible DRAM Address Mapping for SoC-PIM Cooperative On-device LLM Inference},
  author={Seo, Seong Hoon and Kim, Junghoon and Lee, Donghyun and Yoo, Seonah and Moon, Seokwon and Park, Yeonhong and Lee, Jae W},
  booktitle={2025 IEEE International Symposium on High Performance Computer Architecture (HPCA)},
  pages={1720--1733},
  year={2025},
  organization={IEEE}
}

@article{jiang2024mixtral,
  title={Mixtral of experts},
  author={Jiang, Albert Q and Sablayrolles, Alexandre and Roux, Antoine and Mensch, Arthur and Savary, Blanche and Bamford, Chris and Chaplot, Devendra Singh and Casas, Diego de las and Hanna, Emma Bou and Bressand, Florian and others},
  journal={arXiv preprint arXiv:2401.04088},
  year={2024}
}

@article{liu2024deepseek,
  title={Deepseek-v3 technical report},
  author={Liu, Aixin and Feng, Bei and Xue, Bing and Wang, Bingxuan and Wu, Bochao and Lu, Chengda and Zhao, Chenggang and Deng, Chengqi and Zhang, Chenyu and Ruan, Chong and others},
  journal={arXiv preprint arXiv:2412.19437},
  year={2024}
}

@inproceedings{patel2024splitwise,
  title={Splitwise: Efficient generative llm inference using phase splitting},
  author={Patel, Pratyush and Choukse, Esha and Zhang, Chaojie and Shah, Aashaka and Goiri, {\'I}{\~n}igo and Maleki, Saeed and Bianchini, Ricardo},
  booktitle={2024 ACM/IEEE 51st Annual International Symposium on Computer Architecture (ISCA)},
  pages={118--132},
  year={2024},
  organization={IEEE}
}

@inproceedings{fang2025klotski,
  title={Klotski: Efficient Mixture-of-Expert Inference via Expert-Aware Multi-Batch Pipeline},
  author={Fang, Zhiyuan and Huang, Yuegui and Hong, Zicong and Lyu, Yufeng and Chen, Wuhui and Yu, Yue and Yu, Fan and Zheng, Zibin},
  booktitle={Proceedings of the 30th ACM International Conference on Architectural Support for Programming Languages and Operating Systems, Volume 2},
  pages={574--588},
  year={2025}
}

@inproceedings{cao2025moe,
  title={Moe-lightning: High-throughput moe inference on memory-constrained gpus},
  author={Cao, Shiyi and Liu, Shu and Griggs, Tyler and Schafhalter, Peter and Liu, Xiaoxuan and Sheng, Ying and Gonzalez, Joseph E and Zaharia, Matei and Stoica, Ion},
  booktitle={Proceedings of the 30th ACM International Conference on Architectural Support for Programming Languages and Operating Systems, Volume 1},
  pages={715--730},
  year={2025}
}

@misc{vllm-project,
    title = {vLLM: Easy, fast, and cheap LLM serving for everyone.},
    url = {https://github.com/vllm-project/vllm},
    author = {Woosuk Kwon and Zhuohan Li and Siyuan Zhuang and Ying Sheng and Lianmin Zheng and Cody Hao Yu and Joseph E. Gonzalez and Hao Zhang and Ion Stoica.},
    month = {September},
    year = {2023}
}

@inproceedings{kim2024monde,
  title={Monde: Mixture of near-data experts for large-scale sparse models},
  author={Kim, Taehyun and Choi, Kwanseok and Cho, Youngmock and Cho, Jaehoon and Lee, Hyuk-Jae and Sim, Jaewoong},
  booktitle={Proceedings of the 61st ACM/IEEE Design Automation Conference},
  pages={1--6},
  year={2024}
}

@inproceedings{devaux2019true,
  title={The true processing in memory accelerator},
  author={Devaux, Fabrice},
  booktitle={2019 IEEE Hot Chips 31 Symposium (HCS)},
  pages={1--24},
  year={2019},
  organization={IEEE Computer Society}
}

@article{paszke2019pytorch,
  title={Pytorch: An imperative style, high-performance deep learning library},
  author={Paszke, Adam and Gross, Sam and Massa, Francisco and Lerer, Adam and Bradbury, James and Chanan, Gregory and Killeen, Trevor and Lin, Zeming and Gimelshein, Natalia and Antiga, Luca and others},
  journal={Advances in neural information processing systems},
  volume={32},
  year={2019}
}

@misc{cublas,
    title = {cuBLAS Docs},
    autohr={NVIDIA},
    note = {\url{https://docs.nvidia.com/cuda/cublas/index.html}},
    year={2024}
}

@misc{h100paper,
    author= {{NVIDIA}},
    year  = {2023},
    title = {NVIDIA H100 Tensor Core GPU Architecture},
    note  = {\url{https://resources.nvidia.com/en-us-tensor-core/gtc22-whitepaper-hopper
}},
}

@misc{a100paper,
    author= {{NVIDIA}},
    year  = {2020},
    title = {NVIDIA A100 Tensor Core GPU Architecc ture},
    note  = {\url{https://images.nvidia.com/aem-dam/en-zz/Solutions/data-center/nvidia-ampere-architecture-whitepaper.pdf
}},
}

@misc{pulp-solver,
    author= {{Stuart, Mitchell and Anita, Kean and Andrew, Mason and Michael, O'Sullivan and Antony, Phillips and Franco, Peschiera}},
    year  = {2024},
    title = {PulP},
    note  = {\url{https://coin-or.github.io/pulp/}},
}

@inproceedings{dai2024deepseekmoe,
  title={DeepSeekMoE: Towards Ultimate Expert Specialization in Mixture-of-Experts Language Models},
  author={Dai, Damai and Deng, Chengqi and Zhao, Chenggang and Xu, Rx and Gao, Huazuo and Chen, Deli and Li, Jiashi and Zeng, Wangding and Yu, Xingkai and Wu, Y and others},
  booktitle={Proceedings of the 62nd Annual Meeting of the Association for Computational Linguistics (Volume 1: Long Papers)},
  pages={1280--1297},
  year={2024}
}

@misc{gpt-oss,
    author= {{OpenAI}},
    year  = {2025},
    title = {GPT-OSS-120B},
    note  = {\url{https://huggingface.co/openai/gpt-oss-120b}},
}

@misc{ramulator2.0,
    author= {{SAFARI Research Group}},
    year  = {2023},
    title = {Ramulator 2.0},
    note  = {\url{https://github.com/CMU-SAFARI/ramulator2}},
}

@inproceedings{khairy2020accel,
  title={Accel-sim: An extensible simulation framework for validated gpu modeling},
  author={Khairy, Mahmoud and Shen, Zhesheng and Aamodt, Tor M and Rogers, Timothy G},
  booktitle={2020 ACM/IEEE 47th Annual International Symposium on Computer Architecture (ISCA)},
  pages={473--486},
  year={2020},
  organization={IEEE}
}

@inproceedings{bakhoda2009analyzing,
  title={Analyzing CUDA workloads using a detailed GPU simulator},
  author={Bakhoda, Ali and Yuan, George L and Fung, Wilson WL and Wong, Henry and Aamodt, Tor M},
  booktitle={2009 IEEE international symposium on performance analysis of systems and software},
  pages={163--174},
  year={2009},
  organization={IEEE}
}

@misc{synopsys.org,
	howpublished = {\url{http://www.synopsys.com/Tools/ Implementation/RTLSynthesis/DesignCompiler/Pages}},
	title = {Design Compiler},
	author = {Synopsys}
}

@article{clark2016asap7,
  title={ASAP7: A 7-nm finFET predictive process design kit},
  author={Clark, Lawrence T and Vashishtha, Vinay and Shifren, Lucian and Gujja, Aditya and Sinha, Saurabh and Cline, Brian and Ramamurthy, Chandarasekaran and Yeric, Greg},
  journal={Microelectronics Journal},
  volume={53},
  pages={105--115},
  year={2016},
  publisher={Elsevier}
}

@misc{dgx-a100,
    title = {Introduction to the NVIDIA DGX A100 System},
    url = {https://docs.nvidia.com/dgx/dgxa100-user-guide/introduction-to-dgxa100.html},
    author = {NVIDIA},
    year = {2021}
}

@misc{Samsung_HBM2e_3.2Gbps,
  author = {Samsung Semiconductor},
  title = {KHAA44801B-MC16: 8GB HBM2E Flashbolt},
  year = {2020},
  url = {https://semiconductor.samsung.com/dram/hbm/hbm2e-flashbolt/khaa44801b-mc16/}
}

@misc{JEDEC_JESD235D,
  author = {JEDEC Solid State Technology Association},
  title = {High Bandwidth Memory DRAM (HBM1, HBM2) JESD235D},
  year = {2020},
  url = {https://www.jedec.org/sites/default/files/docs/JESD235D.pdf}
}

@article{zhong2025hybrimoe,
  title={HybriMoE: Hybrid CPU-GPU Scheduling and Cache Management for Efficient MoE Inference},
  author={Zhong, Shuzhang and Sun, Yanfan and Liang, Ling and Wang, Runsheng and Huang, Ru and Li, Meng},
  journal={arXiv preprint arXiv:2504.05897},
  year={2025}
}

@misc{Intel_NPU_decode_limited_by_compute,
  author = {{Intel NPU Acceleration Library}},
  title = {Decoding LLM performance—Prefill phase is compute bound on NPU},
  howpublished = {\url{https://intel.github.io/intel-npu-acceleration-library/llm_performance.html}},
  note = {Accessed: 2025-08-18}
}

@article{Dong2025_async_KV_prefetching,
  author = {Yanhao Dong and Yubo Miao and Weinan Li and Xiao Zheng and Chao Wang and Feng Lyu},
  title = {Accelerating LLM Inference Throughput via Asynchronous KV Cache Prefetching},
  journal = {arXiv preprint arXiv:2504.06319},
  year = {2025},
  note = {Accessed: 2025-08-18}
}

@inproceedings{vespa_superpages,
  author    = {Mayank Parasar and Abhishek Bhattacharjee and Tushar Krishna},
  title     = {VESPA: VIPT Enhancements for Superpage Accesses},
  booktitle = {arXiv preprint arXiv:1701.03499},
  year      = {2017}
}

@inproceedings{usenix_atc20_superpage,
  author    = {Weixi Zhu and Alan L. Cox and Scott Rixner},
  title     = {A Comprehensive Analysis of Superpage Management Mechanisms and Policies},
  booktitle = {2020 USENIX Annual Technical Conference (USENIX ATC)},
  year      = {2020}
}

@misc{upmem_sdk,
  author       = {{UPMEM}},
  title        = {UPMEM Software Development Kit (SDK)},
  year         = 2025,
  url          = {https://sdk.upmem.com/}
}

@inproceedings{liao2021ascend,
  title={Ascend: a scalable and unified architecture for ubiquitous deep neural network computing: Industry track paper},
  author={Liao, Heng and Tu, Jiajin and Xia, Jing and Liu, Hu and Zhou, Xiping and Yuan, Honghui and Hu, Yuxing},
  booktitle={2021 IEEE International Symposium on High-Performance Computer Architecture (HPCA)},
  pages={789--801},
  year={2021},
  organization={IEEE}
}

@inproceedings{liao2019davinci,
  title={DaVinci: A scalable architecture for neural network computing},
  author={Liao, Heng and Tu, Jiajin and Xia, Jing and Zhou, Xiping},
  booktitle={2019 IEEE Hot Chips 31 Symposium (HCS)},
  pages={1--44},
  year={2019},
  organization={IEEE Computer Society}
}

@article{norrie2021design,
  title={The design process for Google's training chips: TPUv2 and TPUv3},
  author={Norrie, Thomas and Patil, Nishant and Yoon, Doe Hyun and Kurian, George and Li, Sheng and Laudon, James and Young, Cliff and Jouppi, Norman and Patterson, David},
  journal={IEEE Micro},
  volume={41},
  number={2},
  pages={56--63},
  year={2021},
  publisher={IEEE}
}

@inproceedings{pessl2016drama,
  title={$\{$DRAMA$\}$: Exploiting $\{$DRAM$\}$ addressing for $\{$Cross-CPU$\}$ attacks},
  author={Pessl, Peter and Gruss, Daniel and Maurice, Cl{\'e}mentine and Schwarz, Michael and Mangard, Stefan},
  booktitle={25th USENIX security symposium (USENIX security 16)},
  pages={565--581},
  year={2016}
}

@misc{yu2025orderschaosenhancinglargescale,
      title={Orders in Chaos: Enhancing Large-Scale MoE LLM Serving with Data Movement Forecasting}, 
      author={Zhongkai Yu and Yue Guan and Zihao Yu and Chenyang Zhou and Shuyi Pei and Yangwook Kang and Yufei Ding and Po-An Tsai},
      year={2025},
      archivePrefix={arXiv},
      url={https://arxiv.org/abs/2510.05497}, 
}

@misc{qwen2025qwen15moea27b,
  title={Qwen1.5-MoE-A2.7B}, 
  author={Qwen Team},
  year={2024},
  url={https://huggingface.co/Qwen/Qwen1.5-MoE-A2.7B}
}

@misc{zheng2023lmsyschat1m,
      title={LMSYS-Chat-1M: A Large-Scale Real-World LLM Conversation Dataset}, 
      author={Lianmin Zheng and Wei-Lin Chiang and Ying Sheng and Tianle Li and Siyuan Zhuang and Zhanghao Wu and Yonghao Zhuang and Zhuohan Li and Zi Lin and Eric. P Xing and Joseph E. Gonzalez and Ion Stoica and Hao Zhang},
      year={2023},
      eprint={2309.11998},
      archivePrefix={arXiv},
      primaryClass={cs.CL}
}

@techreport{handy2021future_low_latency_memory,
  title={The Future of Low-Latency Memory: Why Near Memory Requires a New Interface},
  author={Handy, Jim and Coughlin, Tom},
  institution={Objective Analysis and Coughlin Associates},
  year={2021}
}

\end{document}